\documentclass[twoside,twocolumn,9pt]{article}
\usepackage{extsizes}
\usepackage[super,sort&compress,comma]{natbib} 
\usepackage[version=3]{mhchem}
\usepackage[left=1.5cm, right=1.5cm, top=1.785cm, bottom=2.0cm]{geometry}
\usepackage{balance}
\usepackage{mathptmx}
\usepackage{sectsty}
\usepackage{graphicx} 
\usepackage{lastpage}
\usepackage[format=plain,justification=justified,singlelinecheck=false,font={stretch=1.125,small,sf},labelfont=bf,labelsep=space]{caption}
\usepackage{float}
\usepackage{fancyhdr}
\usepackage{fnpos}
\usepackage[english]{babel}
\addto{\captionsenglish}{%
  
}

\usepackage{amsmath,amssymb,amsfonts,amsbsy,amsthm,booktabs,epsfig,graphicx,rotating}
\usepackage{bm}
\usepackage{array}
\usepackage{droidsans}
\usepackage{charter}
\usepackage[T1]{fontenc}
\usepackage[usenames,dvipsnames]{xcolor}
\usepackage{setspace}
\usepackage[compact]{titlesec}
\usepackage{hyperref}

\usepackage{epstopdf}

\definecolor{cream}{RGB}{222,217,201}

\begin{document}

\pagestyle{fancy}
\thispagestyle{plain}
\fancypagestyle{plain}{
\renewcommand{\headrulewidth}{0pt}
}

\makeFNbottom
\makeatletter
\renewcommand\LARGE{\@setfontsize\LARGE{15pt}{17}}
\renewcommand\Large{\@setfontsize\Large{12pt}{14}}
\renewcommand\large{\@setfontsize\large{10pt}{12}}
\renewcommand\footnotesize{\@setfontsize\footnotesize{7pt}{10}}
\makeatother

\renewcommand{\thefootnote}{\fnsymbol{footnote}}
\renewcommand\footnoterule{\vspace*{1pt}%
\color{cream}\hrule width 3.5in height 0.4pt \color{black}\vspace*{5pt}} 
\setcounter{secnumdepth}{5}

\makeatletter 
\renewcommand\@biblabel[1]{#1}            
\renewcommand\@makefntext[1]%
{\noindent\makebox[0pt][r]{\@thefnmark\,}#1}
\makeatother 
\renewcommand{\figurename}{\small{Fig.}~}
\sectionfont{\sffamily\Large}
\subsectionfont{\normalsize}
\subsubsectionfont{\bf}
\setstretch{1.125} 
\setlength{\skip\footins}{0.8cm}
\setlength{\footnotesep}{0.25cm}
\setlength{\jot}{10pt}
\titlespacing*{\section}{0pt}{4pt}{4pt}
\titlespacing*{\subsection}{0pt}{15pt}{1pt}

\fancyfoot{}
\fancyfoot[LO,RE]{\vspace{-7.1pt}\includegraphics[height=9pt]{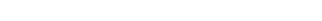}}
\fancyfoot[CO]{\vspace{-7.1pt}\hspace{13.2cm}\includegraphics{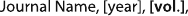}}
\fancyfoot[CE]{\vspace{-7.2pt}\hspace{-14.2cm}\includegraphics{head_foot/RF}}
\fancyfoot[RO]{\footnotesize{\sffamily{1--\pageref{LastPage} ~\textbar  \hspace{2pt}\thepage}}}
\fancyfoot[LE]{\footnotesize{\sffamily{\thepage~\textbar\hspace{3.45cm} 1--\pageref{LastPage}}}}
\fancyhead{}
\renewcommand{\headrulewidth}{0pt} 
\renewcommand{\footrulewidth}{0pt}
\setlength{\arrayrulewidth}{1pt}
\setlength{\columnsep}{6.5mm}
\setlength\bibsep{1pt}

\makeatletter 
\newlength{\figrulesep} 
\setlength{\figrulesep}{0.5\textfloatsep} 

\newcommand{\topfigrule}{\vspace*{-1pt}%
\noindent{\color{cream}\rule[-\figrulesep]{\columnwidth}{1.5pt}} }

\newcommand{\botfigrule}{\vspace*{-2pt}%
\noindent{\color{cream}\rule[\figrulesep]{\columnwidth}{1.5pt}} }

\newcommand{\dblfigrule}{\vspace*{-1pt}%
\noindent{\color{cream}\rule[-\figrulesep]{\textwidth}{1.5pt}} }

\makeatother

\twocolumn[
  \begin{@twocolumnfalse}
{\includegraphics[height=30pt]{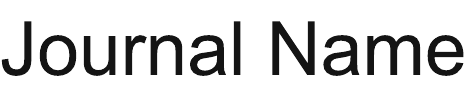}\hfill\raisebox{0pt}[0pt][0pt]{\includegraphics[height=55pt]{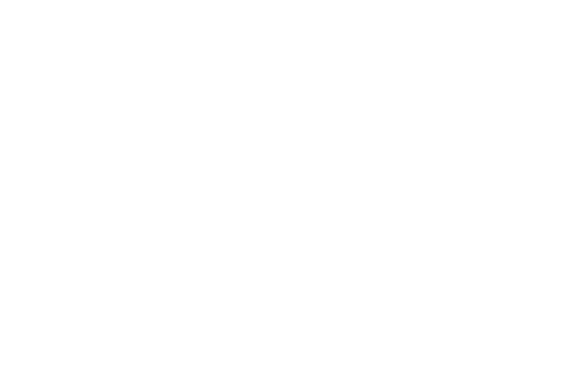}}\\[1ex]
\includegraphics[width=18.5cm]{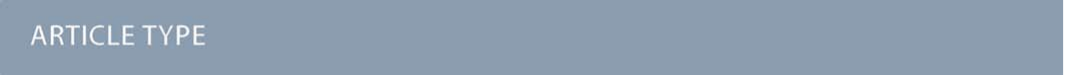}}\par
\vspace{1em}
\sffamily
\begin{tabular}{m{4.5cm} p{13.5cm} }

\includegraphics{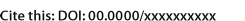} & \noindent\LARGE{\textbf{Spectroscopic constants from atomic properties: a machine learning approach }} \\
\vspace{0.3cm} & \vspace{0.3cm} \\

 & \noindent\large{ Mahmoud A. E. Ibrahim\textit{$^{a,b,c}$}, X. Liu,\textit{$^{d}$} and J. P\'erez-R\'ios \textit{$^{a,b}$}} \\

\includegraphics{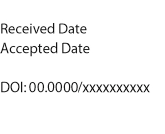} & \noindent\normalsize{We present a machine-learning approach toward predicting spectroscopic constants based on atomic properties. After collecting spectroscopic information on diatomics and generating an extensive database, we employ Gaussian process regression to identify the most efficient characterization of molecules to predict the equilibrium distance, vibrational harmonic frequency, and dissociation energy. As a result, we show that it is possible to predict the equilibrium distance with an absolute error of 0.04~\AA~and vibrational harmonic frequency with an absolute error of 36~cm$^{-1}$, including only atomic properties. These results can be improved by including prior information on molecular properties leading to an absolute error of 0.02~\AA~and 28~cm$^{-1}$ for the equilibrium distance and vibrational harmonic frequency, respectively. In contrast, the dissociation energy is predicted with an absolute error $\lesssim 0.4$~eV. Alongside these results, we prove that it is possible to predict spectroscopic constants of homonuclear molecules from the atomic and molecular properties of heteronuclears. Finally, based on our results, we present a new way to classify diatomic molecules beyond chemical bond properties.
 } \\

\end{tabular}

 \end{@twocolumnfalse} \vspace{0.6cm}

  ]

\renewcommand*\rmdefault{bch}\normalfont\upshape
\rmfamily
\section*{}
\vspace{-1cm}


\footnotetext{\textit{$^{a}$~Department of Physics and Astronomy, Stony Brook University, Stony Brook, New York 11794, USA}}
\footnotetext{\textit{$^{b}$~Institute for Advanced Computational Science, Stony Brook University, Stony Brook, New York 11794, USA}}
\footnotetext{\textit{$^{c}$~Department of Physics, Faculty of Science, Assiut University, Assiut, 71515, Egypt}}
\footnotetext{\textit{$^{d}$~Fritz-Haber-Institut der Max-Planck-Gesellschaft, D-14195 Berlin, Germany}}



\section{\textit{Introduction}}
Since the beginning of molecular spectroscopy in the 1920s, the relationship between spectroscopic constants of diatomic molecules has been an intriguing and captivating matter in chemical physics. Following early attempts by Kratzer, Birge and Mecke \cite{kratzer1920ultraroten,birge1925quantum,mecke1925construction}, Morse proposed a relationship between the equilibrium distance, $R_e$, and the harmonic vibrational frequency, $\omega_e$, as $R_e^2\omega_e=\gamma$, where $\gamma$ is a constant, after analyzing the spectral properties of 16 diatomic molecules\cite{morse1929diatomic}. However, as more spectroscopic data became available, further examination of the Morse relation revealed its applicability to only a tiny number of diatomic molecules\cite{clark1934london}. Next, in a series of papers, Clark et al. generalized Morse's idea via the concept of a periodic table of diatomic molecules. Eventually, Clark's efforts translated into several relations, each limited to specific classes of molecules \cite{clark1934london,clark1934simple,clark1941systematics,clark1941systematicsVI}. Simultaneously, Badger proposed a more neat relationship, including atomic properties of the atoms constituting the molecule\cite{badger1934relation}. Following Badger's proposal, multiple authors have found new relations, which have seen some utility even for polyatomic molecules \cite{kraka2010generalization,kurita2004relationship,badger1935relation}. Nevertheless, Badger's relations are not generalizable to all diatomic molecules~\cite{kaupp2017chemistry,cioslowski2000badger,penney1936relation}. In general, several empirical relationships between $R_e$ and $\omega_e$ were proposed in the  1930s  and the 1940s \cite{allen1935internuclear,huggins1935molecular,huggins1936molecular,sutherland1938relation,linnett1940relation,clark1941systematicsVI,clark1941systematics,wu1944relation,wu1947relation,guggenheimer1946new,linnett1945force,gordy1946relation}. In summary, from 1920 till now, the number of empirical relations published is around 70 collected by Kraka et al. \cite{kraka2010generalization}. Most of these empirical relations were tested by several authors, finding some constraints on their applicability \cite{liu2021relationship,kraka2010generalization,kaupp2017chemistry,penney1936relation,cioslowski2000badger}. However, all of these relationships were based on empirical evidence rather than on a given physical or chemical principle. 

On the other hand, in 1939 Newing proposed a theoretical justification for observed empirical relationships between spectroscopic constants given by 
\begin{equation} \label{eq:1}
   cf(R_e)=\mu \omega_e^2
\end{equation}
where $c$ is a constant for \textit{similar} molecules, $f(R_e)$ is some function of the equilibrium distance, and $\mu$ is the reduced mass of the molecule. In particular, Newing used Slater's application of the virial theorem, concluding that the empirical laws may be related to the existence of a \emph{universal} repulsive field in diatomic molecules~\cite{slater1933virial,newing1940xxx}. The theoretical justification given by Newing implies that several relations of the form given by (\ref{eq:1}) exist, each of which holds for a set of \textit{similar} diatomic molecules; however, for any practical application of these empirical laws, the sets of \textit{similar} diatomic molecules must be identified first. The approach was not viable because similarity needs to be defined precisely. 

In the 1960s and 1970s, a number of authors devised the virial theorem, perturbation theory, and Helmann-Feynman theorem \cite{guttinger1932verhalten,pauli1933principles,hellmann1937einfuhrung,feynman1939forces} to develop a better understanding of the nature of the relationship between $R_e$ and $\omega_e$ via electron densities \cite{salem1963theoretical,empedocles1967force,empedocles1968force,anderson1969relationships,anderson1970vibrational,anderson1971universal,simons1971development,anderson1972effective}. Most notably, Anderson and Parr were able to establish a relationship between $R_e$, $\omega_e$, and atomic numbers $Z_1$ and $Z_2$, as
\begin{equation} \label{eq:2}
    \omega_e=\sqrt{\frac{4 \pi A Z_1 Z_2 \exp(-\xi R_e)}{\mu}},
\end{equation}
where $\mu$ is the reduced mass of the molecule, and $A$ and $\xi$ (the electron density decay constant) are fitting parameters. Further, assuming that $R_e$ is given by the sum of atomic radii of the constituent atoms and following simple arguments using the electron density function, it can be shown that
\begin{equation} \label{eq:3}
    R_e=\frac{1}{\xi'}\ln\left(\frac{Z_1Z_2}{B}\right),
\end{equation}
where it is assumed that the electron density has a given decay constant $\xi'$, and $B$ is a fitting parameter. Using Eqs.~(\ref{eq:2}) and (\ref{eq:3}), one finds 
\begin{equation} \label{eq:4}
    \omega_e=\sqrt{\frac{C (Z_1Z_2)^{-\eta}}{\mu}},
\end{equation}
where $C=4\pi AB^{(1+\eta)}$ and $\eta=(\xi'-\xi)/\xi'$. Anderson and Parr found that taking $C$, $\xi$ and $\xi'$ as functions of the groups and periods of the constituent atoms results in better fitting~\cite{anderson1971universal,anderson1972effective}.  Anderson and Par tested their relationships against 186 molecules and agreed reasonably with experimental values. Recently Liu et al. tested Eqs.~(\ref{eq:2}) and (\ref{eq:3}) against an extended data set of 256 molecules, finding that these relationships lead to errors $\gtrsim 10\%$ upon adding more data \cite{liu2021relationship}. Therefore, these relationships are not universal and further study is required to elucidate proper relationships. However, the pioneering work of Anderson, Parr, and others provided well-motivated relationships between spectroscopic constants theoretically for the first time. Most significantly, their work pointed towards a possible direct connection between a diatomic molecule's spectroscopic properties and its individual atoms' atomic properties and positions in the periodic table. 

Alongside these developments, several authors attempted connecting the dissociation energy, $D_0^0$, with $\omega_e$ and $R_e$ of diatomic molecules\cite{sutherland1938relation,sutherland1940determination,somayajulu1960dissociation,lippincott1961dissociation,gazquez1979universal,wiener1997relationships}. However, these received little attention due to the lack of reliable experimental data~\cite{badger1934relation,sutherland1940determination,skinner1954dissociation,jhung1990universal,luo2007comprehensive}. Most of the relationships are given by
\begin{equation} \label{eq:5}
    D_0^0=A' \mu \omega_e^2 R_e^l
\end{equation}
where $A'$ and $l$ are constants depending on the form and parameterization of the potential energy functions that describe the molecule. For instance, Sutherland found that by taking $A'$ as a function of groups and periods, better results can be obtained \cite{sutherland1938relation,sutherland1940determination}.

Thanks to machine learning (ML) techniques and the development of extensive spectroscopic databases~\cite{Liu2020}, it has been possible to study the relationship between spectroscopic constants from a heuristic perspective, i.e., from a data-driven approach~\cite{liu2021relationship}, find optimal potentials based on spectroscopy data~\cite{Stevenson2019} and to improve {\it ab initio} potentials to match experimental observations~\cite{Fu2022}. In particular, Gaussian process regression (GPR) models have been used on a large dataset of 256 heteronuclear diatomic molecules. As a result, it was possible to predict $R_e$ from the atomic properties of the constituent atoms. Similarly, $\omega_e$ and the binding energy $D_0^0$ were predicted using combinations of atomic and molecular properties. However, the work of Liu et al. only studied heteronuclear molecules. Hence, the universality of the relationship between spectroscopic constants still needs to be revised.
On the other hand, ML techniques can be used to boost density functional theory approaches to larger systems with low computational costs~\cite{comp1,comp2,comp3,comp4}. Hence, ML techniques are used to enlarge the capabilities of quantum chemistry methods. However, if sufficient data and information are available, could ML substitute quantum chemistry methods?

In this work, we present a novel study on the relationship between spectroscopic constants via ML models, including homonuclear molecules in a dataset of 339 molecules: the largest dataset of diatomics ever used. As a result, first, we show that it is possible to predict $R_e$ and $\omega_e$ with mean absolute errors $\sim$ 0.026~\AA~and $\sim$26 cm$^{-1}$, leading to an improvement of factor 2 in predicted power and accuracy concerning previous ML models. Furthermore, the dissociation energy, $D_0^0$, is predicted with a mean absolute error $\sim$0.4 eV, in accordance with previous ML models. However, our model can benefit from having a more accurate and extensive database. Second, we show that it is possible to accurately predict the molecular properties of homonuclear molecules out of heteronuclear ones. Finally, since we use the same ML model in this work, we are in a unique situation to define similarity among molecules. Thus, we propose a data-driven classification of molecules. The paper is organized as follows: in Section 2, we introduce the database and analyze the main properties; in Section 3, we present the ML models with a particular emphasis on Gaussian process regression; in Section 4, we present our results and in Section 5, the conclusions.

\section{The data set}

In this work, we extend the data set by Liu et al. \cite{liu2021relationship,Liu2020} by adding the ground state spectroscopic constants of 32 homonuclear and 54 extra heteronuclear diatomic molecules from Refs.~\cite{huber2013molecular,4AuF,5AuF,6AuF,7AuF,8AuF,NeWAuF,9CaD,reCoF,10CoF,11CoF,12CoF,13CoF,14CoF,15CoF,16CoO,17CoO,18CoO,19CrF,20InBr,21CrC,22IrSi,23LiBe,24LiBe,25LiCa,26LiCa,27LiCa,LiCanew,28LiCs,29LiCs,30LiCs,31LiK,32LiK,33LiNaKKLiLiNa,34LiMg,35LiNa,36MgCa,37MgCa,38MgHCaHSrHBaH,39MoCRuCPdCNbCZrC,40MoC,41Na2NaKK2,42NaRb,43NaRb,44NaRb,45NbC,46Ne2Ar2Kr2Xe2,47NiC,48NiO,49NiS,50PbISnI,51PdC,52SrI,53SrI,54SrI,55SrI,56SrI,57SrI,58TiF,59TiF,60TiF,61ZnBr,62UF,63VF,UO,65YC,66YC,67ZrC,3ArNeArXe,ArXetheory1, ArXetheory2}. The dataset counts 339 entries based on experimentally determined spectroscopic constants: $R_e$ is available for 339, $\omega_e$ for 327, and D$_0^0$ is available for 249 molecules. 

\begin{figure*}[t]
\begin{center}
 \includegraphics[width=1\linewidth]{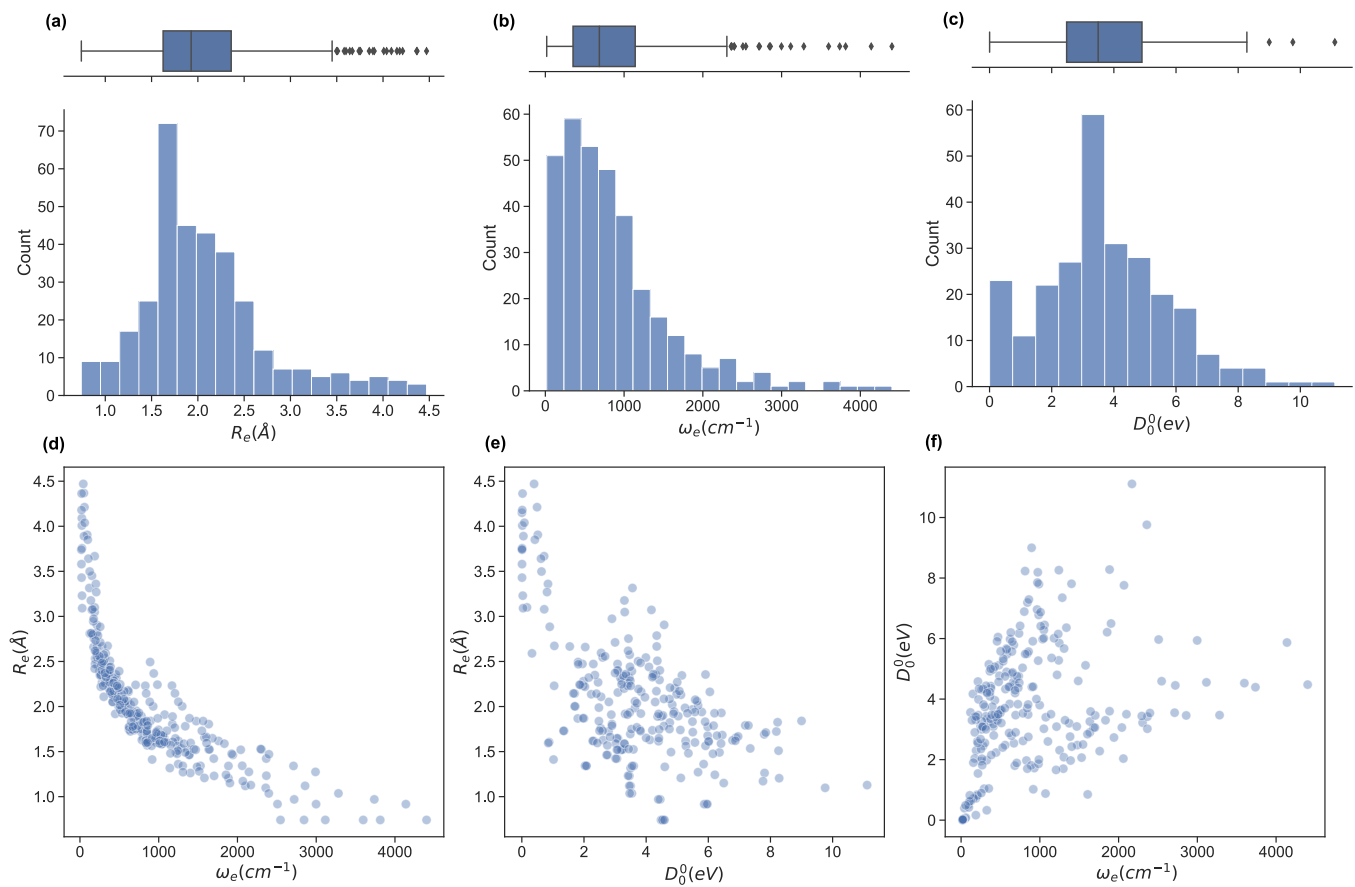}  
\caption{\label{fig:hist} The dataset of diatomic molecules' ground state spectroscopic constants. Panels (a)-(c) display the distribution of the main spectroscopic constants in the dataset--$R_e$, $\omega_e$ and $D_{0}^{0}$-- via a histogram representation and a box plot (at the top) for each. Panels (d)-(f) show the relationship between different spectroscopic constants of the molecules in the database.}
\end{center}
\end{figure*}

\begin{figure}[h]
\begin{center}
 \includegraphics[width=1\linewidth]{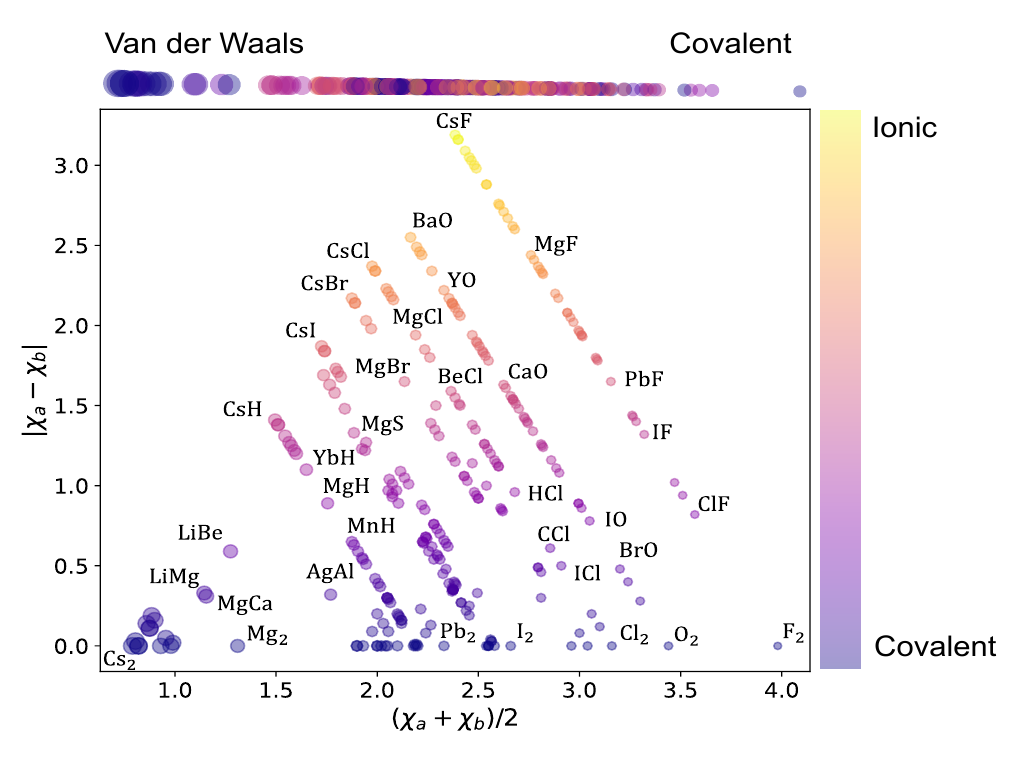}  
\caption{\label{fig:tri}  Arkel-Ketelaar's triangle of the dataset. The average electronegativity of the constituent atoms on the x-axis, the difference in electronegativity of the constituent atoms correlates with the ionic character on the y-axis. The color of each circle demonstrates the ionic character of the corresponding molecule following the color bar on the right of the figure. The size of the circles differentiates between covalent (smaller circles) and van der Waals (larger circles) molecules, as illustrated at the top of the figure.}
\end{center}
\end{figure}

To assess the variation of the spectroscopic constants in the dataset, we display the histogram and box plots of $R_e$, $\omega_e$, and $D_{0}^{0}$ in Fig.~\ref{fig:hist}. This Figure shows that the spectroscopic constants' histogram is nearly unimodal. However, $R_e$ and $\omega_e$ show a heavy-tailed distribution. In the case of $R_e$, the tail is due to the presence of van der Waals molecules. In contrast, light molecules are responsible for the tail in the histogram for $\omega_e$. On the other hand, the box plot of $D_{0}^{0}$ shows almost no outliers and only an incipient peak for a molecule with binding energy smaller than 0.75 eV due to the presence of van der Waals molecules. On the other hand, we investigate the relationship between pairs of spectroscopic constants in panels (d)-(f) of Fig.~\ref{fig:hist}. For example, panel (d) displays $R_e$ versus $\omega_e$, showing an exponential trend similar to the one suggested by Eq.~(\ref{eq:2}) or a power law (Morse relationship). On the contrary, by plotting $R_e$ versus $D_0^0$ and  $D_0^0$ versus $\omega_e$  in panels (e) and (f), respectively, we notice a large dispersion of the points with no observed trends in both panels. 

Next, we analyze the chemical properties of the molecules under consideration, employing the Arkel-Ketelaar triangle--also known as the Jensen triangle, which separates qualitatively covalent, ionic, and van der Waals molecules. The triangle plots the absolute value of the electronegativity difference between the constituent atoms $|\chi_a-\chi_b|$ versus their average electronegativity $\left(\chi_a+\chi_b\right)/2$, as shown in Fig.~\ref{fig:tri}, where $\chi_a$ and $\chi_b$ denote the electronegativities of the molecules' constituent atoms. The average electronegativity of the constituent atoms on the x-axis quantifies the van der Waals-covalent bonding. On the contrary, the difference in electronegativity of the constituent atoms quantifies the ionic character on the y-axis. The triangle shows that the data set comprises chemically diverse diatomic molecules with bonding characters ranging from covalent to ionic with many van der Walls. This chemical diversity strongly manifests itself in the range of the ground state spectroscopic constants depicted in Fig.~\ref{fig:hist}.  

\section{The machine learning (ML) model}

Machine learning (ML) is a vast discipline that utilizes data-driven algorithms to perform specific tasks (e.g., classification, regression, clustering). Among the different ML techniques, in this work, we use Gaussian process regression (GPR), which is especially suitable for small datasets. This section briefly describes GPR and our methods to generate and evaluate models.

\subsection{Gaussian process regression} 

We define our data set $D=\{(\textbf{x}_i,y_i)|i=1,\cdot\cdot\cdot,n\}$, where $\textbf{x}_i$ is a feature vector of some dimension $D$ associated with the $i$-th element of the dataset, $y_i$ is a scalar target label, and $n$ is the number of observations, i.e., the number of elements in the dataset. The set of all feature vectors and corresponding labels can be grouped in the random variables $X$ and $\textbf{y}$, respectively, where $X=(\textbf{x}_1,\cdot\cdot\cdot,\textbf{x}_n)$ and $\textbf{y}=(y_1,\cdot\cdot\cdot,y_n)$. Here, $\textbf{y}$ consists of values of molecular properties to be learned. $y_i$ is $R_e$, $\omega_e$, or $D_0^0$ of the $i$-th molecule, whereas $\textbf{x}_i$ is a vector containing atomic or molecular properties of the same molecule.

We are interested in mapping features to target labels via a regression model $y_i=f(\textbf{x}_i)+\varepsilon_i$, where $f(\textbf{x}_i)$ is the regression function, and $\varepsilon_i$ is an additive noise. We further assume that $\varepsilon_i$ follows an independent, identically distributed (i.i.d) Gaussian distribution with variance $\sigma_n^2$
\begin{equation} \label{noise}
\bm{\varepsilon} \sim \mathcal{N}(0,\sigma_y^2I)
\end{equation}  
where $\bm{\varepsilon}=(\varepsilon_1,\cdot\cdot\cdot,\varepsilon_n)$ and $I$ is the identity matrix. 

One approach to tackle the regression problem is to specify a functional form of $f(\textbf{x}_i)$. Then, set the free parameters of the regression model by fitting the data. Alternatively, one can disregard specifying a functional form of $f(\textbf{x}_i)$ but instead place a prior distribution over a space of functions and infer the posterior predictive distribution following a Bayesian non-parametric approach. Within this group of methods, we find Gaussian process regression (GPR), assuming a Gaussian process prior $\mathcal{GP}$ over the space of functions \cite{10.7551/mitpress/3206.001.0001,pedregosa2011scikit}.
\begin{equation}
    f(\textbf{x}) \sim \mathcal{G P} \left(m(\textbf{x}),k(\textbf{x},\textbf{x}')\right).
\end{equation}
A Gaussian process is a collection of random variables, any finite number of which have a joint Gaussian distribution. A Gaussian process is specified by a mean function $m(\textbf{x})$ and a covariance function(kernel) $k(\textbf{x},\textbf{x}')$, we will describe both shortly. A posterior distribution of the value of $f(\textbf{x}^*)$ at some point of interest, $\textbf{x}^*$, is determined through the Bayes theorem as 

\begin{equation}
    f(\textbf{x}^*)|X,Y \sim \mathcal{N}(\mu^*,\Sigma^*)
\end{equation}
where
\begin{equation}
\label{eq6}
   \begin{gathered}  
   \mu^*=m(\textbf{x}^*)+k(\textbf{x}^*,X)\left[k(X,X)+\sigma_{n}^{2}I\right]^{-1}(\textbf{y}-m(X)). \\
   \Sigma^*=k(\textbf{x}^*,X)\left[k(X,X)+\sigma_{n}^{2}I\right]k(X,\textbf{x}^*).
   \end{gathered} 
\end{equation}
The mean of the resulting predictive posterior distribution, $\mu^*$, is used to obtain a point estimate of the value of $f(\textbf{x}^*)$, and its covariance $\Sigma^*$ provides a confidence interval. 

In GPR, the regression model is completely specified by the kernel $k(\textbf{x},\textbf{x}')$.  The kernel is a similarity measure that specifies the correlation between a pair of values $f(\textbf{x})$ and $f(\textbf{x}')$ by only using the distance between a pair of feature vectors $\textbf{x}$ and $\textbf{x}'$ as its input variable. Specifying a kernel, we encode high-level structural assumptions (e.g., smoothness, periodicity, etc.) about the regression function. Here, we focus on the Mat\'ern class kernel given by
 given by
\begin{equation}
\label{eq10}
    k(\textbf{x}_p,\textbf{x}_q)=\sigma_f^2\frac{2^{1-\nu}}{\Gamma(\nu)}\frac{\sqrt{2}{d(\textbf{x}_p,\textbf{x}_q)}^{\nu}}{l}K_{\nu}\left(\frac{\sqrt{2}{d(\textbf{x}_p,\textbf{x}_q)}^{\nu}}{l}\right) + \sigma_n^2 \delta_{pq},
\end{equation}
where $d(\textbf{x}_p,\textbf{x}_q)$ is the Euclidean distance between $\textbf{x}$ and $\textbf{x}'$, $K_{\nu}(z)$ is the modified Bessel function of the second kind of order $\nu$ and argument $z$, $\Gamma(x)$ is the Euler gamma function of argument $x$, $l$ is the characteristic length scale, $\sigma_f^2$ is the signal variance, and $\delta_{pq}$ is the Kronecker delta. $\nu$ controls the smoothness of the process. For instance, for $\nu=1/2$, the process is zero times differentiable. On the contrary, the process is infinitely differentiable at the limit $\nu \rightarrow \infty$: the so-called radial basis function (RBF) kernel. Values of $\nu$  that are suitable for regression applications are 1/2, 3/2, 5/2, and $\infty$ \cite{10.7551/mitpress/3206.001.0001}.

We can encode a physical model via the relationships between spectroscopic constants by specifying the Gaussian process prior mean function $m(\textbf{x})$. A common choice of the prior mean function is $m(\textbf{x})=0$. This choice is satisfactory in most cases, especially in interpolation tasks. However, selecting an appropriate prior mean function can simplify the learning process (delivering better results using fewer data). The mean function can also guide the model for better predictions as $k(\textbf{x}_p,\textbf{x}_q) \rightarrow 0$; this is necessary for extrapolation and interpolation among sparse data points. Further, a model with a specified mean function is more interpretable.     

\subsection{Model development and performance evaluation}
\label{Modeld and e}

Its parameters and hyperparameters characterize a GPR model. Parameters involve $(\sigma_n,l,\sigma_f)$ of Eq.~(\ref{eq10}) plus additional parameters depending on the form of the prior mean function. On the contrary, hyperparameters involve selecting features, the form of the prior mean function, and the order $\nu$ of the Mat\'ern kernel. To determine the parameters and the hyperparameters, we divide the dataset $D$ into two subsets: $D_{\text{tv}}$ and $D_{\text{test}}$. First, $D_{\text{tv}}$ is used for the training and validation stage, in which we determine the model's hyperparameters. Then, $D_{\text{test}}$, known as the test set, is left out for model final testing and evaluation and does not take any part in determining the parameters nor the hyperparameters of the model. 

To design a model, we choose an $X$ suitable to learn $\textbf{y}$ through a GPR. We then choose a convenient prior mean function $m(X)$ based on physical intuition, alongside the last hyperparameter $\nu \in \{ 1/2,3/2,5/2,\infty \}$ is determined by running four models, each with a possible value of $\nu$, and we chose the one that performs the best on the training data to be the final model. Precisely, a cross-validation (CV) scheme is used to evaluate the performance of each model iteratively: we split $D_{\text{tv}}$ into a training set $D_{\text{train}}$ ($\sim 90 \%$ of $D_{\text{tv}}$) and a validation set $D_{\text{valid}}$. We use $D_{\text{train}}$ to fit the model and determine its parameters by maximizing the log-marginal likelihood. The fitted model is then used to predict the target labels of $D_{\text{valid}}$. We repeat the process with a different split in each iteration such that each element in $D_{\text{tv}}$ has been sampled at least once in both $D_{\text{train}}$ and $D_{\text{valid}}$. After many iterations, we can determine the average performance of the model. We compare the average performance of the four models after the CV process. Finally, We determine the value of $\nu$ to be its value for the best-performing model. 

We adopt a Monte Carlo (MC) splitting scheme to generate the CV splits. Using the MC splitting scheme, we expose the models to various data compositions, and thus, we can make more confident judgments about our models' performance and generality ~\cite{liu2021relationship}. To generate a single split, we use stratified sampling\cite{pedregosa2011scikit,botev2017variance}. First, we stratify the training set into smaller strata based on the target label. Stratification will be such that molecules in each stratum have values within some lower and upper bounds of the target label (spectroscopic constant) of interest. Then, we sample the validation set so that each stratum is represented. Stratified sampling minimizes the change in the proportions of the data set composition upon MC splitting, ensuring that the trained model can make predictions over the full range of the target variable. Using the Monte Carlo splitting scheme with cross-validation (MC-CV) allows our models to train on $D_{\text{tv}}$ in full, as well as make predictions for each molecule in $D_{\text{tv}}$. In each iteration, $D_{\text{valid}}$ simulates the testing set; thus, by the end of the MC-CV process, it provides an evaluation of the model performance against $\sim$90\% of the molecules in the data set before the final testing stage.  

We use 1000 MC-CV iterations to evaluate each model's performance. Two estimators evaluate the models' performance at each iteration, the mean absolute error (MAE) and the root mean squared error (RMSE), given by
\begin{equation}
   \begin{gathered}  
   \text{MAE}= \frac{1}{N} \sum_{i=1}^{N} |y_i-y_i^*|, \\
   \text{RMSE}= \sqrt{\frac{1}{N} \sum_{i=1}^{N} (y_i-y_i^*)^2}
   \end{gathered} 
\end{equation}
where $y_i^*$ and $y_i$ are the true and predicted values, respectively, and $N$ is the number of observations in consideration. We report the final training/validation $\overline{\text{MAE}}$ and $\overline{\text{RMSE}}$ with the sample standard deviation (STD) and the standard error of the means (SEM) given by
\begin{equation}
   \begin{gathered}  
   \overline{\text{MAE}}=\frac{1}{M} \sum_{i=1}^{M} \text{MAE}_i, \\
   \overline{\text{RMSE}}= \frac{1}{M} \sum_{i=1}^{M} \text{RMSE}_i
   \end{gathered} 
\end{equation}
Where $M$ is the number of the MC-CV iterations, and MAE$_i$ (RMSE$_i$) is the MAE (RMSE) of the $i$th MC-CV iteration.
\begin{equation}
   \begin{gathered}  
   \text{STD}({\text{MAE}})= \sqrt{\frac{1}{M-1}\sum_{i=1}^{M}( \overline{\text{MAE}}-\text{MAE}_i)^2}, \\
   \text{STD}(\text{RMSE})=\sqrt{ \frac{1}{M-1} \sum_{i=1}^{M} (\overline{\text{RMSE}}-\text{RMSE}_i)^2}
   \end{gathered} 
\end{equation}
\begin{equation}
   \begin{gathered}
\text{SEM(MAE)}=\frac{\text{STD}({\text{MAE}})}{\sqrt{M}}, \\  
\text{SEM(RMSE)}=\frac{\text{STD}({\text{RMSE}})}{\sqrt{M}}   
   \end{gathered} 
\end{equation}
We use learning curves to evaluate the performance of the models as a function of the size of $D_{\text{train}}$. We use 500 CV splits at each training set size to generate the learning curves. The validation and training  $\overline{\text{RMSE}}$ $\pm$ 0.5STD(RMSE) are plotted as a function of the size of $D_{\text{train}}$. 

Models that have the lowest validation $\overline{\text{MAE}}$, $\overline{\text{RMSE}}$, and SEM are elected for the testing stage. In the testing stage, we fit the models to $D_{\text{tv}}$ and make predictions of the target labels of $D_{\text{test}}$. Finally, we report the validation $\overline{\text{MAE}} \pm \text{SEM(MAE)}$ and $\overline{\text{RMSE}} \pm \text{SEM(RMSE)}$ and test MAE and RMSE as our final evaluation of the model. 

\section{Results and discussion}

We have developed seven new models to predict $R_e$, $\omega_e$, and $D_0^0$: r2, r3, and r4 to predict $R_e$, models for predicting $\omega_e$ are denoted w2, w3, and w4, while only one model is used to predict $D_0^0$, labeled as d1. In addition, we implemented two of the best-performing models of Liu et al.~\cite{liu2021relationship} (denoted r1 and w1) using our updated dataset and compared them with our models. All models are divided into three categories: (\textit{i}) r1, r2, and w2 only employ atomic properties as features in the kernel and as variables in the prior mean function, (\textit{ii}) r3 and w3 employ atomic properties as features in the kernel but use spectroscopic data in the prior mean function, and (\textit{iii}) r4, w4, and d1 include spectroscopic data both in the kernel and the prior mean function. 

\begin{table*}[h]
\begin{center}
\caption{Machine learning models summary. The target column includes the molecular property to be predicted. The model column refers to the ML model used. The molecules column refers to the number of molecules in the training plus validation set $D_{\text{tv}}$. Features are the atomic and molecular properties employed to characterize every data point in the kernel. Prior mean stands for the prior mean function used for each model as indicated in the text, and $\nu$ represents the order of the Mat\'ern kernel determined via the MC-CV scheme described in section \ref{Modeld and e}. }
\begin{tabular*}{\textwidth}{@{\extracolsep{\fill}}lcclcc}
Target & Model &   Molecules  & Features & Prior mean  &  $\nu$  \\
\hline
\\
$R_e$(\AA)& rlr1 &  314 & $p_1 + p_2, \ g_1 + g_2, \ \ln{(Z_1 Z_2)}$ & - & 
-
\\ \\
& rlr2 &  314 & $\ln{(\omega_e)}, \ p_1 + p_2, \ g_1 + g_2, \ \ln{(Z_1 Z_2)},\ \ln{(\mu)}$ & - & 
-
\\ \\
& r1 &  314 & $p_1, g_1, p_2,g_2$ & $m_{r1}$& 
1/2
\\ \\

   & r2 & $314$ &$p_1$, $g_1$, $p_2$, $g_2$, $\mu^{1/2}$ &$m_{r2}$  & $3/2$ \\ \\
  
  & r3  & $308$ &$p_1,g_1,p_2,g_2,\mu^{1/2}$ &$m_{r3}$ & $3/2$ \\ \\
  & r4  & $310$ &$\ln(\omega_e)$, $p_1,g_1,p_2,g_2,\mu^{1/2}$ &$m_{r3}$ & $3/2$ \\ \\
 \\
 $\ln(\omega_e)$ & wlr1 &  $308$ & $p_1+p_2, \ g_{1}+g_{2},\ \ln{(Z_1 Z_2)},\ \ln{(\mu)}$ & - & - \\ \\
 & wlr2 &  $308$ & $R_e,\ p_1+p_2, \ g_{1}+g_{2}, \ \ln{(Z_1 Z_2)},\ \ln{(\mu)}$ & - & - \\ \\
& w1 &  $308$ & $R_{e}^{-1},p_1, g_{1}^{iso},p_2, g_{2}^{iso}, \bar{g}$& 0 & $5/2 $\\ \\ 
 & w2 &  $308$ & $p_1$, $g_1$, $p_2$, $g_2$, $\mu^{1/2}$& $m_{w2}$& $3/2 $\\ \\
 & w3 &  $308$ & $p_1$, $g_1$, $p_2$, $g_2$, $\mu^{1/2}$& $m_{w3}$ & $3/2 $\\ \\
 & w4 &  $308$ & $R_e$,$p_1$, $g_1$, $p_2$, $g_2$, $\mu^{1/2}$& $m_{w4}$  & $3/2$ \\ \\ 
  \\
 $\ln(D_0^0)$ & dlr1 &  $244$ & $\ln{(R_e)}$, $\ln{(\omega_e)}$, $p_1+p_2$, $g_1+g_2$, $\ln{(\mu)}$& - & -
 \\  \\
 & d1 &  $244$ & $p_1$, $g_1$, $p_2$, $g_2$, $\mu^{1/2}$ & $m_{d1}$ & 
3/2
\label{table:1}
\end{tabular*}
\end{center}
\end{table*}

In all the seven newly developed models, we use the groups $g_1$ and $g_2$ and periods $p_1$ and $p_2$ of the molecules' constituent atoms and the square root of the reduced mass of the molecule $\mu^{1/2}$ as features. Therefore, the set of properties \{ $p_1$,$g_1$,$p_2$,$g_2$,$\mu^{1/2}$\} uniquely defines each molecule in the dataset. On the contrary, additional spectroscopic properties are added to these five features for models within the category (\textit{iii}). Furthermore, we choose the models' features and prior mean functions using physical intuition based on the discussion in the introduction and observations from the data Fig.~\ref{fig:hist}, and $\nu$ was set to 3/2 using the CV scheme discussed in the last section. The models' characteristics are given in Table \ref{table:1}.

For all the nine implemented models, we permute the groups and periods in $D_{\text{train}}$ in the training and validation stage and in $D_{\text{tv}}$ in the testing stage to impose permutational invariance \cite{liu2021relationship}. That is, the models should not differentiate between $\textbf{x}=(p_1,g_1,p_2,g_2,...)$ and $\textbf{x}'=(p_2,g_2,p_1,g_1,...)$ upon exchanging the numbering of the two atoms in a molecule. Eight of the models use linear prior mean functions, the linear coefficients of which are determined by fitting the linear model to $D_{\text{train}}$ in each CV iteration in the training and validation stage and by fitting to $D_{\text{tv}}$ in the testing stage.

For the sake of comparison with baseline ML models we have implemented linear regression (LR) models based on equations (\ref{eq:3}-\ref{eq:5}). Specifically, models rlr1 and rlr2 to predict $R_e$, wlr1 and wlr2 to predict $\ln({\omega_e})$ and dlr1 to predict $\ln(D_O^O)$. The same MC-CV scheme used to train the GPR models was used to train the LR models. A description of these models is given in table \ref{table:1} and a statistical summary of their performance is given in table \ref{table:2}.

\subsection{$R_e$}

The first spectroscopic constant under consideration is the equilibrium distance, $R_e$. We have implemented and developed two models for predicting $R_e$ using only atomic properties: r1 and r2, as detailed in Table~\ref{table:1}. Model r1 is the same as in Liu et al.~\cite{liu2021relationship} using groups and periods of the constituent atoms as features. The model r2 requires an extra feature related to the reduced mass of the molecule. For both models, we explicitly express the models' prior mean functions as linear functions in the groups and periods of the diatomic molecules' constituent atoms.
\begin{equation}
   m_{r1-r2} = \beta_0^{r1-r2}+\beta_1^{r1-r2}(p_1+p_2) + \beta_2^{r1-r2}(g_1+g_2),
\end{equation}
where  $\beta_k^{r1-r2}$, $k \in \{0,1,2\}$ are the linear coefficients of  $m_{r1-r2}$.

\begin{figure*}[h]
\begin{center}
 \includegraphics[width=\linewidth]{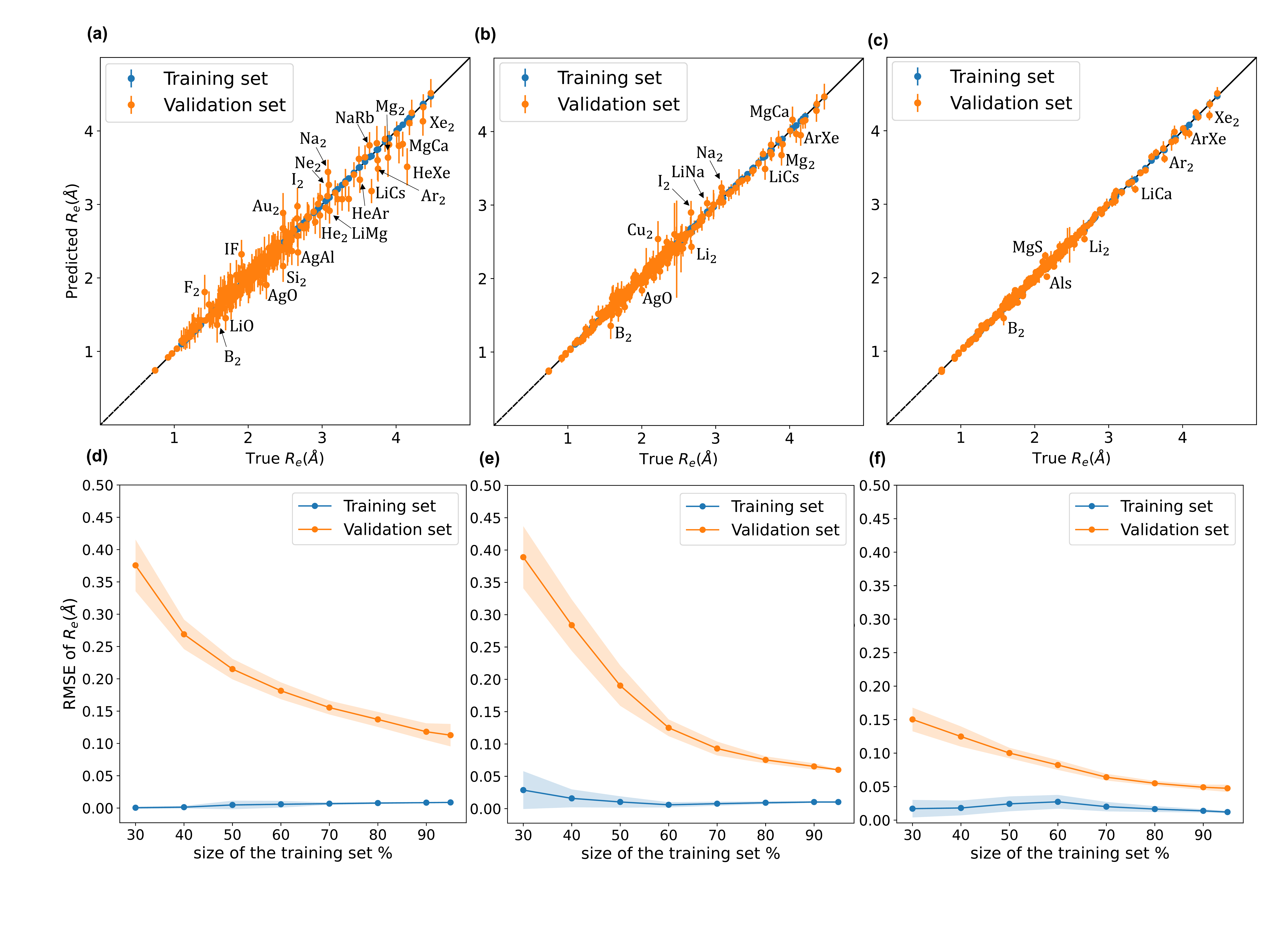}  
\caption{\label{fig:Re scatter}  Upper row show scatter plots of experimental values of $R_e$ on the x-axis and predicted $R_e$ on the y-axis via models (a) r1 (b) r2  (c) r3, points, and error bars represent the predictive distribution means and standard deviations respectively after averaging over 1000 MC-CV steps. The lower row shows three learning curves of models (d) r1, (e) r2, and (f) r3. Points  and shade around represent the $\overline{\text{RMSE}}$ and $\pm$ 0.5 $\text{STD}(\overline{\text{RMSE}})$ over 500 MC-CV splits.}
\end{center}
\end{figure*}

\begin{table*}[h]
\begin{center}
\caption{Statistical summary of the performance of the GPR models using different features, kernels, and prior mean functions as listed in Table \ref{table:1}.  The performance of each model is evaluated using both the validation and test scores. The values for MAE and RMSE with $^*$ show an SEM $\lesssim$ 0.001 \AA. }
  \begin{tabular*}{\textwidth}{@{\extracolsep{\fill}}lccccc}
 Target& Model &\begin{tabular}{@{}cc@{}}
                 Validation\\
                  $\overline{\text{MAE}}$ $\pm$ SEM
                 \end{tabular}  & \begin{tabular}{@{}cc@{}}
                 Validation\\
                   $\overline{\text{RMSE}}$ $\pm$ SEM
                 \end{tabular}&\begin{tabular}{@{}cc@{}}
                 Test\\
                   MAE
                 \end{tabular}& \begin{tabular}{@{}cc@{}}
                 Test\\
                   RMSE
                 \end{tabular}\\
\hline
\\
$R_e$(\AA) &  rlr1 & 0.265 & 0.422 & - & - \\ \\
&  rlr2 & 0.109 & 0.143 & - & - \\ \\
&  r1 & 0.060$^*$ & 0.100$^*$&0.051 &0.077 \\ \\

   & r2 & 0.041$^*$  & 0.060$^*$  & 0.044& 0.064\\ \\
  
  & r3  & 0.027$^*$ & 0.039$^*$ & 0.026&0.038  \\ \\
  & r4  & 0.026$^*$ & 0.038$^*$ & 0.025&0.041 \\ \\
 \\ $\omega_e$(cm$^{-1}$) &  wlr1 & 238 & 338 & - & - \\ \\
&  wlr2 & 134 & 227 & - & - \\ \\
&  w1 &$33.2$ $\pm$ $0.3$ &$64.8$ $\pm \ 1.0$& 34.1& 62.6\\ \\
 & w2 &$40.3$ $\pm$ $0.3$ &$66.3$ $\pm \ 0.6$& 36.2&49.97 \\ \\
 & w3 &27.7 $\pm$ 0.2 &44.8 $\pm \ 0.4$&33.97 & 45.14\\ \\
 
 & w4 &25.9 $\pm$ 0.2 &41.6 $\pm \ 0.3$& 28.8 & 39.9\\ \\ 
 \\
$D_0^0$ (eV) & dlr1 & 0.99 & 1.25 & - & - \\ \\ 
$\ln{(D_0^0)}$ & d1 & 0.12 $\pm$ 0.04 & 0.17 $\pm$  $0.05$&0.13 & 0.18 \\ 
\label{table:2}
\end{tabular*}
\end{center}
\end{table*}
A comparison between models r1 and r2 is displayed in Fig.~\ref{fig:Re scatter}. The scatter plots show a more significant dispersion of the predictions for model r1 compared to model r2. Both models show the same outliers: homonuclear and van der Waals molecules. However, for model r2, the number of outliers is smaller than in model r1, and their dispersion from the true line is significantly suppressed. As a result, model r2 performs substantially better, especially in predicting molecules with $R_e\geq$ 3~\AA (mainly van der Waals molecules). The learning curves of models r1 and r2, displayed in panels (d) and (e) of Fig.~\ref{fig:Re scatter}, respectively, show a convergent validation curve towards the training set result as the size of the training set increases, indicative of the learning capability of the model, although, model r2 displays a faster convergence, indicating that the model learns more efficiently. Overall, model r2 shows an improvement in the prediction on $R_e$ $\sim 20\%$ with respect r1 as shown in Table\ref{table:2}, leading to validation $\overline{\text{MAE}}$ and $\overline{\text{RMSE}}$ of 0.041~\AA~and 0.060~\AA~respectively.

Motivated by previously proposed relationships between $R_e$ and $\ln{(\omega_e)}$, we introduce models r3 and r4. Model r3 employs the same features as model r2 but incorporates spectroscopic information in the prior mean function. On the contrary, model r4 uses $\ln{(\omega_e)}$ as a feature. Both models have a prior mean given by
\begin{equation}
    \begin{gathered} 
         m_{r3-r4} = \beta_0^{r3-r4}+\beta_1^{r3-r4}(p_1+p_2) + \beta_2^{r3-r4}(g_1+g_2) \\ + \beta_3^{r3-r4} \ln{(\mu^{1/2}})+\beta_4^{r3-r4} \ln{(\omega_e)}, 
    \end{gathered}
\end{equation}
where $\beta_k^{r3-r4}$, $k \in \{0,1,2,3,4\}$ are linear coefficients of $m_{r3-r4}$. The two models have similar performance as shown in table~\ref{table:2}. The results of model r3 are presented in panels (c) and (f) of Fig.~\ref{fig:Re scatter}. Panel (c) shows a minimal scatter around the true line. The error bars are suppressed compared with panels (a) and (b) of the same figure, indicating a higher confidence level of the model's predictions. The validation curve in Panel (f) shows that the learning rate of model r3 is significantly higher than the other two models. Using 50\% of the available data is sufficient for model r3 to achieve a validation $\overline{\text{RMSE}}$ comparable to model r1 using 90$\%$ of the data set. Overall, models r3 and r4 show an improvement in the prediction on $R_e$ $\sim 40\%$ as shown in Table\ref{table:2}, leading to validation $\overline{\text{MAE}}$ of 0.027~\AA~and 0.026~\AA~and a validation $\overline{\text{RMSE}}$ of~0.039~\AA~ and 0.038~\AA~ respectively. In other words, models r3 and r4 are almost two times more precise in predicting $R_e$ than previously ML-based or empirically-derived predictions, and almost as accurate as the state-of-the-art {\it ab initio} calculations for diatomics~\cite{Liu2023,ladjimi2023diatomic}. Furthermore, the lower panes of Fig.~\ref{fig:Re scatter} show converging learning curves characterized by relatively narrow gaps between validation and training curves. The decaying trend of the validation curves suggests that convergence toward lower levels of errors is possible upon further training on more extensive datasets. The training MAE of r2 is $\sim  6 \times 10^{-3}$~\AA~; this means that we might have the capacity to achieve an accuracy $\sim 0.010$~\AA~  using only atomic properties. In other words, with more data our ML models could be as accurate as {\it ab initio} quantum chemistry methods.   

To highlight a few of the common outliers of the four models, we consider Li$_2$, B$_2$, LiCs, and LiCa. r1, r2, r3, r4 underestimate $R_e$ for Li$_2$ by 6 \% - 10 \%,.  r1, r2, r3, r4 underestimate $R_e$ for B$_2$ by 14\%. 15\%, 7\%, and 8\%, respectively, which could be connected to B$_2$ being the only homonuclear molecule from group 13 in the data set. For LiCs, we have found $R_e=3.67$~\AA~\cite{30LiCs} r2 predicts $R_e=3.49\pm 0.15$~\AA; that is, the experimental value is 1.2 standard deviation away from the mean predictive posterior distribution of model r2 for LiCs, most of the theoretical $R_e$ values of LiCs are within one standard deviation~\cite{29LiCs}. For LiCa, the experimental value found by Krois et al. is $R_e=3.34$~\AA~\cite{LiCanew}. On the contrary, the r4 model predicts $R_e=3.20 \pm 0.05$~\AA~, almost three standard deviations away from the experimental value. However, model r2 predicts $R_e=3.33 \pm 0.09$~\AA~, with only 0.3\% relative error. In addition, high-level {\it ab initio} calculations results are within one standard deviation from the mean predictive posterior distribution of model r2 for LiCa, namely CASPT2 $R_e=3.40$~\AA~\cite{LiCath2}, QCISD(T) $R_e=3.41$~\AA~\cite{LiCath1}, MRCI  $R_e=3.40$~\AA~\cite{LiCath1}, and CIPI $R_e=3.40$~\AA~\cite{LiCath3}.

\subsection{$\omega_e$}
We have implemented and developed four models to predict $\omega_e$ as listed in Table~\ref{table:1}. Model w1 is the best-performing model of Liu et al.~\cite{liu2021relationship}. It is characterized by six features, including atomic and molecular properties. Namely, the groups and periods of the constituent atoms, the average group, $\bar{g}=(g^{iso}_1+g^{iso}_2)/2$, and $R_e^{-1}$. $g^{iso}$ encodes isotopic information, such that $g_i^{iso}=0$ for deuterium, $g_i^{iso}=-1$ for tritium, and $g_i^{iso}=g_i$ for every other element. The prior mean function is set to zero. On the other hand, model w2 only includes groups and periods of the constituent atoms and the reduced mass of the molecule. The prior mean of model w2 is given by 
\begin{equation}
    m_{w2} =\beta_0^{w2}+\beta_1^{w2}(p_1+p_2) + \beta_2^{w2}(g_1+g_2)+\beta_3^{w2} \ln{(\mu^{1/2})},
\end{equation}
where $\beta_k^{w2-w3}$, $k \in \{0,1,2,3\}$ are the linear coefficients of $m_{w2}$. Model w3 uses the same features as model w2 but includes $R_e$ in the prior mean function. Model w4 is characterized by six features as model w1 and uses $R_e$ as a feature in both the kernel and in the prior mean function. 

Motivated by the relationship between $\omega_e$ and $R_e$, both w3 and w4 use the same prior mean function
\begin{equation}
\label{eq15}
    \begin{gathered}  
        m_{w3-w4} = \beta_0^{w3-w4}+\beta_1^{w3-w4}(p_1+p_2) + \beta_2^{w3-w4}(g_1+g_2) \\ + \beta_3^{w3-w4} R_e +\beta_4^{w3-w4} \ln{(\mu^{1/2})},
    \end{gathered}  
\end{equation}
where $\beta_k^{w4}$, $k \in \{0,1,2,3,4\}$ are the linear coefficients of $m_{w3-w4}$.The inclusion of the reduced mass in models w2, w3, and w4 eliminates the necessity of imposing isotopic information on the groups of constituent atoms.

\begin{figure*}[h]
\begin{center}
 \includegraphics[width=1\linewidth]{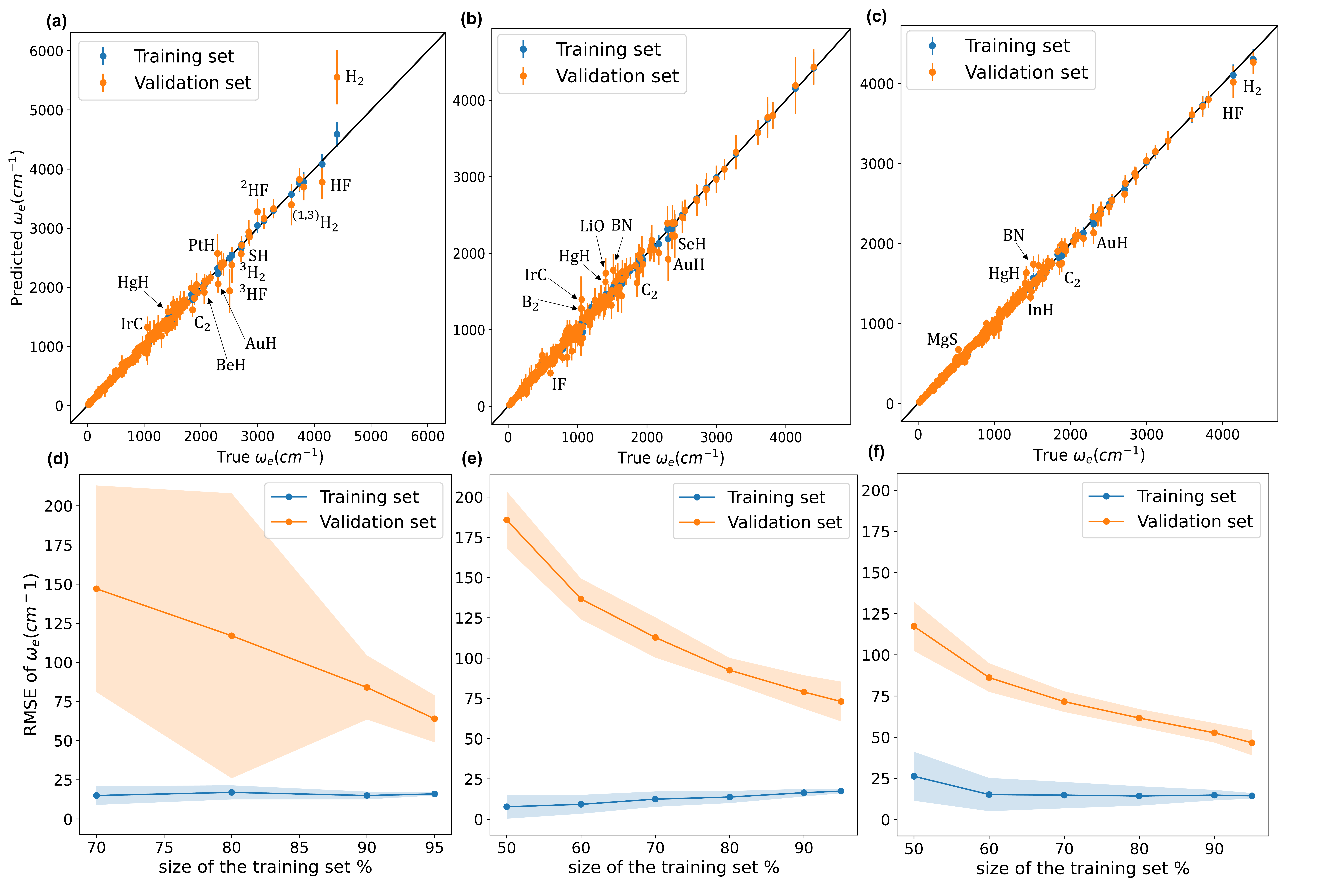}  
\caption{\label{We scatter}  Upper row show scatter plots of experimental values of $\omega_e$ on the x-axis and predicted $\omega_e$ on the y-axis via models (a) w1 (b) w2  (c) w4, points, and error bars represent the predictive distribution means and standard deviations respectively after averaging over 1000 MC-CV steps. The lower row shows three learning curves of models (d) w1, (e) w2, and (f) w4. Points  and shade around represent the $\overline{\text{RMSE}}$ and $\pm$ 0.5 $\text{STD}(\overline{\text{RMSE}})$ over 500 MC-CV splits.}
\end{center}
\end{figure*}

Fig.~\ref{We scatter} compares w1, w2, and w4 (plots of w3 are similar to those of w4). We notice from panel (a) that model w1 struggles against hydrides, and hydrogen and hydrogen fluoride isotopologues. Indeed, the model significantly overestimates $\omega_e$ for H$_2$. On the other hand, panel (b) shows that w2 performs much better against hydrides, and hydrogen and hydrogen fluoride isotopologues. w2 predictions for H$_2$ and HF are accurate and even better than those of models w3 and w4, as shown in panel (c). Panels (a) and (b) clearly show that model w2 outperforms model w1 when considering molecules with larger values of $\omega_e$. Looking at the learning curves in panels (d) and (e), we see that model w2 is far more consistent than model w3, as indicated by the shade around the validation curves of both models. From Table \ref{table:2}, the validation SEM(RMSE) of models w2 and w1 show that model w2 is 40 \% more consistent in its performance than model w1 when both models are validated using the same 1000 MC-CV splits. Furthermore, the test RMSE of w2 is 20 \%  lower than that of w1. Model w2 has lower dimensionality than model w1 and only implements atomic properties; nevertheless, it performs similarly to model w1. 

From Table \ref{table:2}, we see that although model w3 has a test MAE almost equal to model w1, models w3 and w4 have validation $\overline{\text{MAE}}$s  15 \% - 21 \% lower than that of w1, indicating an overall better average performance of the newly developed models. Furthermore, w3 and w4 have validation $\overline{\text{RMSE}}$s  and test RMSEs 28 \% - 36 \% lower than w1, showing the robustness of the two new models. Panel (c) of Fig.~\ref{We scatter} shows minimal scatter around the true line. Few hydrides, along with BN and C$_2$, still challenge the model; however, their absolute errors are significantly suppressed compared to w1 and w2. The validation curve of model w4 in panel (f) shows a much higher learning rate than w1 and w2, with a much shallower gap between the validation and learning curves. Moreover, The shadow around the validation curve is minimal at all training sizes. From table \ref{table:2}, we see that w3 and w4 are far more consistent than w1, with STD(RMSE) 60 \% - 70 \% lower than that of w1. 

On the other hand, the lower three panels in Fig.~\ref{We scatter} show that the validation and training curves can converge towards lower error values. Hence, all the models might benefit from training on a more extensive dataset. The training MAEs of w1, w2, w3, and w4 range between 8 to 7 cm$^{-1}$, so it might be possible to reach near spectroscopic accuracy ($\sim 10$ cm$^{-1}$) by training these models on larger datasets. In the case of w2, if the validation curve's decaying trend persists upon further training, near spectroscopic accuracy might be achieved only through knowledge of atomic positions in the periodic table. Similarly, these models trained in larger database could outperform the state-of-the-art {\it ab initio} quantum chemistry methods~\cite{Liu2023,ladjimi2023diatomic}.

We highlight some of the outliers that are common to some of the models. All the models overestimate $\omega_e$ for HgH by at least 12 \%, while for IrC, w1 and w2 overestimate $\omega_e$ by 30 \% and 25 \%, while w3 and w4 only overestimate it by only 4 \% and 7 \%, respectively. The observed overestimation might be because HgH and IrC are the only molecules that consist of mercury or iridium in the dataset. 

We have found two values of $\omega_e$ for AuF in the literature; Saenger et al. reported $\omega_e=560$ cm$^{-1}$ in 1992  \cite{NeWAuF}, while Andreev et al. reported $\omega_e=448$ cm$^{-1}$ in 2000 \cite{5AuF}. All our models predict values closer to 560 cm$^{-1}$: w2 predicts $\omega_e=529 \pm 87$ cm$^{-1}$, while w3 and w4 are almost in exact agreement with Saenger's value with $\omega_e=568 \pm 54$ cm$^{-1}$ and $\omega_e=565 \pm 45$ cm$^{-1}$, respectively.  Our predictions are in agreement with relativistic density functional and {\it ab initio} methods. Namely, first-order relativistic density functional calculation predicts $\omega_e=491$ cm$^{-1}$ while Zeroth-order regular relativistic approximation within the Kohn-Sham density functional scheme ZORA(MP) predicts $\omega_e=526$ cm$^{-1}$  \cite{7AuF}. In the same line, the relativistic MP2 approach predicts $\omega_e=590$ cm$^{-1}$ \cite{AuFMP2}, while relativistic MR-CI predicts $\omega_e=525$ cm$^{-1}$ \cite{AuFMRCI}. A similar situation occurs in the case of ZnBr, as shown in table \ref{table:test}. For 30 years, there was a discrepancy in the value of $\omega_e$ of ZnBr. Gosavi et al. reported $\omega_e\approx 319$ cm$^{-1}$ in 1971 \cite{ZnBr319}. Next, Givan et al. reported $\omega_e\approx 198$ cm$^{-1}$ in 1982\cite{ZnBr198}. on the contrary, the MRCI calculations by Elmoussaoui and Korek predicted $\omega_e \approx 267$ cm$^{-1}$ in 2015 \cite{ZnBrMRCI}. Finally, Burton et al. experimentally reported $\omega_e = 284$ cm$^{-1}$ in 2019\cite{61ZnBr}. Here, w2 predicts $\omega_e = 271.2 \pm 21.7$~cm$^{-1}$, w3 predicts $\omega_e = 289.5 \pm 15.4$~cm$^{-1}$ and w4 predicts $\omega_e = 281.0 \pm 12.0$~cm$^{-1}$. Therefore, our predicted values are in great agreement with the most recent theoretical and experimental values.

\begin{table*}[p]
\small
\begin{center}
\caption{Predictions and experimental values of $R_e$ and $\omega_e$ for 25 molecules in the testing set. References of experimental values are included. Ref. column includes references for experimental values}
\begin{tabular*}{\textwidth}{@{\extracolsep{\fill}}lcccccc}
 Molecule&\begin{tabular}{@{}cc@{}}
                 Models \\
                  for \\
                 $R_e$, $\omega_e$\\
                 \end{tabular} &\begin{tabular}{@{}cc@{}}
                 Predicted \\
                 $R_e$(\AA)\\
                 \end{tabular}&\begin{tabular}{@{}cc@{}}
                 Experimental \\
                 $R_e$(\AA)\\
                 \end{tabular}&\begin{tabular}{@{}cc@{}}
                 Predicted \\
                 $\omega_e$ (cm$^{-1}$)\\
                 \end{tabular}&\begin{tabular}{@{}cc@{}}
                 Experimental \\
                 $\omega_e$ (cm$^{-1}$)\\
                 \end{tabular}&Ref.\\
\hline
\\
HCl& r4,w4 &1.267 $\pm$ 0.031&1.274& 2934 $\pm$ 126&2990&\cite{huber2013molecular}\\ 
\\
& r2,w2 &1.267 $\pm$ 0.046&& 3019 $\pm$ 211&&\\ 
\\
$^2$HCl & r4,w4   &1.267 $\pm$ 0.031&1.274& 2173 $\pm$ 89.0&2145&\cite{huber2013molecular}\\ 
\\
&r2,w2   &1.267 $\pm$ 0.0461&& 2122 $\pm$137&&\\ 
\\
RuC& r4,w4  &1.615 $\pm$ 0.043&1.600&1102 $\pm$ 65.0 &1100&\cite{39MoCRuCPdCNbCZrC}\\
\\
&r2,w2  &1.64 $\pm$ 0.075&&1062 $\pm$ 119 &&\\
\\
WO & r4,w4  &1.668 $\pm$ 0.050&1.657& 1044 $\pm$ 72.0 &1067&\cite{WO},\cite{WO2}\\
\\
&r2,w2  &1.70 $\pm$ 0.089&&992.9 $\pm$ 131 &&\\
\\
MoC & r4,w4   &1.650 $\pm$ 0.040&1.676&$981$ $\pm$ $54.0$ &1008&\cite{39MoCRuCPdCNbCZrC},\cite{40MoC}\\
\\
&r2,w2  &1.71 $\pm$ 0.057&&1009 $\pm$ 106 &&\\
\\
 WC& r4,w4  &1.754$\pm$ 0.059 &1.714&$1090$ $\pm$ $87.0$ &983.0&\cite{WC}\\ 
 \\
 &r2,w2  &1.65 $\pm$ 0.099&&1122 $\pm$ 178 &&\\
\\
NbC& r4,w4  & 1.721$\pm$ 0.043 &1.700&985 $\pm$ 58.0 &980.0&\cite{45NbC}\\ 
\\
&r2,w2  &1.665$\pm$ 0.057&&943.0 $\pm$ 107 &&\\
\\
NiC& r4,w4  &1.607 $\pm$ 0.053 &1.627&840 $\pm$ 60.0 &875.0&\cite{47NiC}\\
\\
&r2,w2  &1.667$\pm$ 0.093&&806.1 $\pm$ 111 &&\\
\\
PdC& r4,w4  & 1.740 $\pm$ 0.034 &1.712&876 $\pm$ 42.0 &847.0&\cite{51PdC}\\
  \\ 
&r2,w2  &1.721 $\pm$ 0.058&&873.7 $\pm$ 75.0 &&\\
\\
UO& r4,w4  &1.862 $\pm$ 0.024 &1.838&887 $\pm$ 30.0 &846.0&\cite{UO}\\ 
 \\ 
 &r2,w2  &1.839 $\pm$ 0.03&&893 $\pm$ 45.8 &&\\
\\
NiO&r4,w4  &1.587 $\pm$ 0.041 &1.627&790 $\pm$ 45.0 &839.0&\cite{48NiO}\\ 
 \\ 
 &r2,w2  &1.665 $\pm$ 0.056&&805.6 $\pm$ 84.1 &&\\
\\
\label{table:test}
\end{tabular*}
\end{center}
\end{table*}

\begin{table*}[p]
\ContinuedFloat
\small
\begin{center}
\caption{Continued}
\begin{tabular*}{\textwidth}{@{\extracolsep{\fill}}lcccccc}
 Molecule&\begin{tabular}{@{}cc@{}}
                 Models \\
                  for \\
                 $R_e$, $\omega_e$\\
                 \end{tabular} &\begin{tabular}{@{}cc@{}}
                 Predicted \\
                 $R_e$(\AA)\\
                 \end{tabular}&\begin{tabular}{@{}cc@{}}
                 Experimental \\
                 $R_e$(\AA)\\
                 \end{tabular}&\begin{tabular}{@{}cc@{}}
                 Predicted \\
                 $\omega_e$ (cm$^{-1}$)\\
                 \end{tabular}&\begin{tabular}{@{}cc@{}}
                 Experimental \\
                 $\omega_e$ (cm$^{-1}$)\\
                 \end{tabular}&Ref.\\
\hline
\\
YC&r4,w4  & 1.887	$\pm$ 0.080 &2.050&630 $\pm$ 74.0 &686.0$\pm$ 20&\cite{65YC},\cite{66YC}\\ 
\\ 
ZnF& r4,w4  & 1.765  $\pm$ 0.031&1.768&603 $\pm$ 27.0 &631.0&\cite{ZnF}\\
\\ 
&r2,w2  &1.800$\pm$ 0.054&&582.5 $\pm$ 46.2 &&\\
\\
NiS& r4,w4  & 1.957 $\pm$ 0.048 &1.962&490 $\pm$ 32.0 &512.0&\cite{49NiS}\\
\\ 
&r2,w2  &2.009 $\pm$ 0.082&&484.7 $\pm$ 59.5 &&\\
\\
ZnCl& r4,w4  & 2.135 $\pm$  0.031&2.130&377 $\pm$ 16.0 &390.0&\cite{ZnCl}\\
\\ 
&r2,w2  &2.160 $\pm$ 0.053&&372.1 $\pm$ 29.3 &&\\
\\
ZnBr& r4,w4  & 2.297$\pm$ 0.032 &2.268&281 $\pm$ 12.0 &284.0&\cite{61ZnBr}\\
& &-&-&-&319.0&\cite{ZnBr319}\\
& &-&-&- &198.0&\cite{ZnBr198}\\
\\ 
&r2,w2  &2.316 $\pm$ 0.0542&&271.2 $\pm$ 21.7 &&\\
\\
ZnI& r4,w4  &2.492 $\pm$ 0.033 &2.460&234 $\pm$ 11.0 &223.0&\cite{huber2013molecular}\\
\\ 
&r2,w2  &2.482 $\pm$ 0.0575&&228.1 $\pm$ 19.3 &&\\
\\
InBr& r4,w4  & 2.535 $\pm$ 0.036 &2.543&221 $\pm$ 11.0 &221.0&\cite{20InBr}\\
\\ 
&r2,w2  &2.541 $\pm$ 0.0645&&221.8 $\pm$ 19.8 &&\\
\\
SnI& r4,w4  &2.736 $\pm$	0.038 &2.732&195 $\pm$ 10.0 &197.0&\cite{50PbISnI}\\ 
\\ 
&r2,w2  &2.727 $\pm$ 0.0685&&198.0 $\pm$ 19.9 &&\\
\\
PbI& r4,w4  &2.797 $\pm$ 0.032 &2.798&159 $\pm$ 7.00 &160.0&\cite{50PbISnI}\\
\\ 
&r2,w2  &2.815 $\pm$ 0.0571&&154.1 $\pm$ 13.0 &&\\
\\
CrC&r2 &1.52 $\pm$ 0.0982 &1.630&-&-&\cite{21CrC}\\
\\ 
CoO&r2 &1.543 $\pm$ 0.0566 &1.628&-&-&\cite{16CoO}, \cite{17CoO}, \cite{18CoO}\\
\\ 
IrSi&r2 &2.109 $\pm$ 0.1723 &2.09&-&-&\cite{22IrSi}\\
\\ 
UF&r2 &1.999 $\pm$ 0.082&2.02&-&-&\cite{62UF}\\
\\ 
ZrC&r2 &1.842 $\pm$ 0.0581&1.740&-&-&\cite{67ZrC}\\
\\ 
\label{table:test}
\end{tabular*}
\end{center}
\end{table*}

\subsection{$D_0^0$}

\begin{figure}[h]
\begin{center}
 \includegraphics[width=1\linewidth]{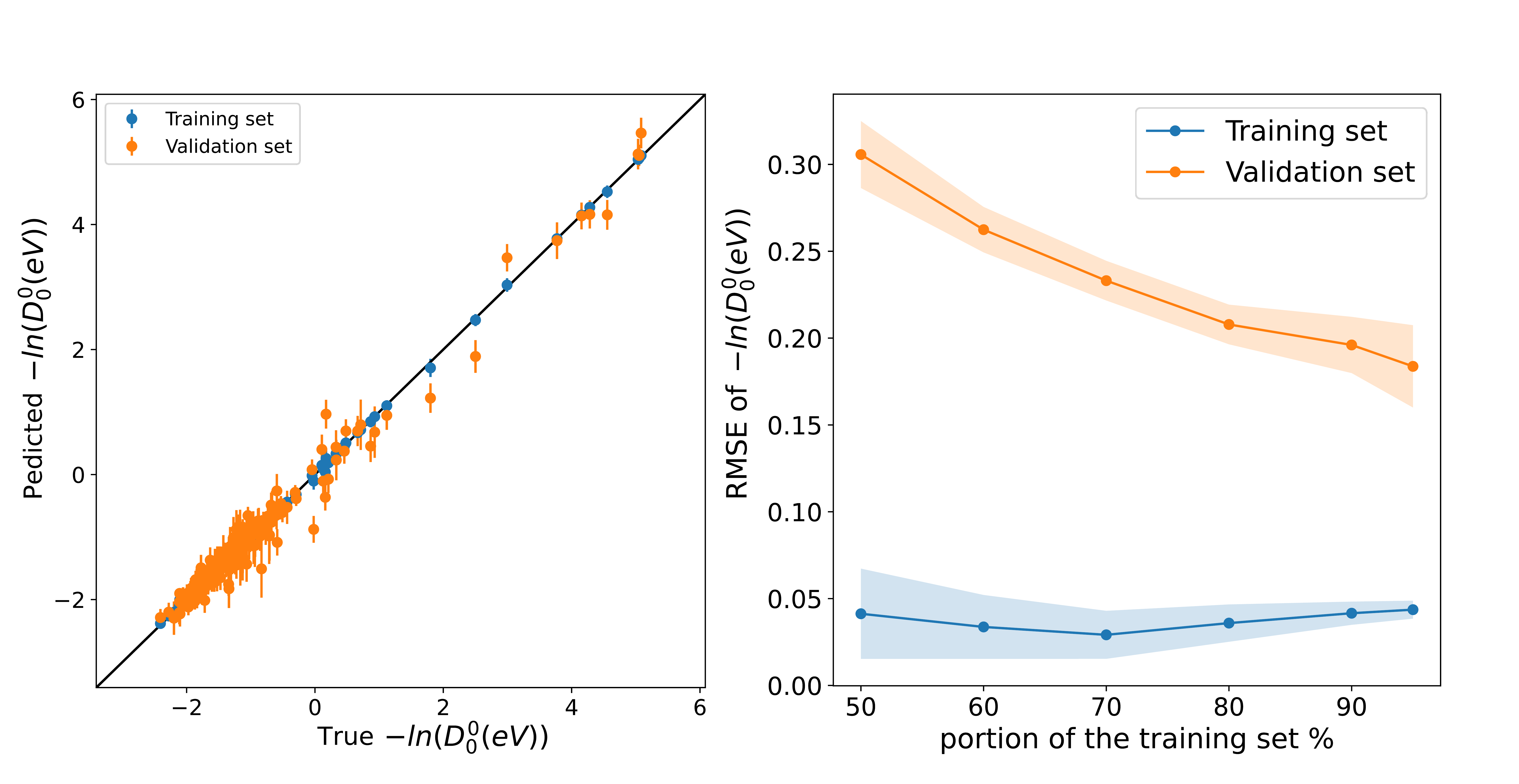}  
\caption{\label{fig: -ln(D)}(a) Scatter plot of experimental values of $-\ln(D_0^0)$ on the x-axis and predicted $-\ln(D_0^0)$ on the y-axis via models d1, points, and error bars represent the predictive distribution means and standard deviations respectively after averaging over 1000 MC-CV steps. (b) shows the learning curves of model d1. Points  and shade around represent the $\overline{\text{RMSE}}$ and $\pm$ 0.5 $\text{STD}(\overline{\text{RMSE}})$ over 500 MC-CV splits.}
\end{center}
\end{figure}

Finally, we have developed model d1 to predict the dissociation energy, $D_0^0$, via $\ln(D_0^0)$ using  ($p_1$,$g_1$,$p_2$,$g_2$,$\mu^{1/2}$) as features in a Mat\'ern 3/2 kernel, and a prior mean function that employs both $\omega_e$ and $R_e$
\begin{eqnarray}
    m_{d1} = \beta_0^{d1}+\beta_1^{d1}(p_1+p_2) + \beta_2^{d1}(g_1+g_2) + \beta_3^{d1} R_e \nonumber \\
    + \beta_4^{d1} \ln{(\mu^{1/2})}+ \beta_5^{d1} \ln{(\omega_e)},
\end{eqnarray}
where $\beta_k^{d1}$, $k \in \{0,1,2,3,4,5\}$ are the linear coefficients of $m_{d1}$. The performance of the model is displayed in Fig.~\ref{fig: -ln(D)}, where the scatter plot [panel (a)] shows some dispersion of the model predictions concerning the true values. From Table~\ref{table:2}, the validation and test errors suggest that the model is consistent and generalizable to new data indicating that model d1 yields reasonable performance as far as $\ln(D_0^0)$ is concerned. However, converting $\ln(D_0^0)$ back to $D_0^0$, the errors are $\sim 0.4$~eV$\equiv 10 \ \text{kcal/mol}$, which is a significant error considering the typical chemical accuracy ($\pm 1 \ \text{kcal/mol}$). However, as shown in panel (b) of Fig.~\ref {fig: -ln(D)}, the model might benefit from training on more data, leading to a potential improvement of a factor of 3. On the other hand, it is possible to accurately predict bond energies, in complex molecules, by using intuitive chemical descriptors, as shown in Refs.\cite{Qu2013,Raza2019}, which is something that we are planning on implementing in the future.

During the development of this work, we have realized that, historically, uncertainties about the dissociation energy experimental values had restrained the development of empirical relations connecting them to other atomic and molecular properties and have led several authors to focus their efforts on the $\omega_e$ - $R_e$ relation \cite{badger1934relation, sutherland1940determination,jhung1990universal}. More recently, Fu et al. used an ML model to predict dissociation energies for diatomic molecules, exploiting microscopic and macroscopic properties~\cite{fu2020joint}. They tested their model against CO and highlighted that the reported experimental dissociation energy in the literature had increased by 100 kcal/mol over the course of 78 years from 1936 to 2014 \cite{fu2020joint,vol1955structure,kepa2014aangstrom} (in Table 1 of Ref.~\cite{fu2020joint}). The data used to train model d1 is primarily collected from Huber and Herzberg's constants of diatomic molecules, first published in 1979 \cite{huber2013molecular}. Unlike experimental values of $R_e$ and $\omega_e$, since 1980, a significant number of $D_0^0$ values have been updated \cite{luo2007comprehensive}. To name a few, MgD, MgBr, MgO, CaCl. CaO, SrI, SrO, TiS, NbO, AgF, AgBr, and BrF all have their experimental values updated with at least  $\pm 2.3 \ \text{kcal/mol}$ difference from their values in Huber and Herzberg \cite{huber2013molecular,luo2007comprehensive}. Moreover, for some molecules, the uncertainties in $D_0^0$ experimental values are not within chemical accuracy. For instance, MgH, CaCl, CaO, CaS, SrH, BaO, BaS, ScF, Tif, NbO, and BrF have uncertainties ranging from $\pm 1 \ \text{kcal/mol} \ \text{up to} \pm 8 \  \text{kcal/mol}$ \cite{luo2007comprehensive}.

Based on the previous discussion, we can connect the unsatisfactory performance of model d1--in comparison to our developed $R_e$ and $\omega_e$ models--to noise modeling. Unlike $R_e$ and $\omega_e$, it is most likely that uncertainties around $D_0^0$ experimental values drive from various systematic effects. Therefore, modeling the errors in $D_0^0$ experimental values to be identically distributed as in Eq.~(\ref{noise}) might not be a proper treatment. Thus, to develop better models for predicting $D_0^0$, more sophisticated techniques of error modeling might be required. To this endeavor, gathering more reliable data with experimental uncertainty within $\pm 1 \ \text{kcal/mol}$ might be sufficient. Something that we are working on it, and it will be published elsewhere.

\begin{table*}[h]
\begin{center}
\caption{Predictions via model d1 and experimental values of $D_0^0$ for seven molecules in the testing set. References of experimental values are included. Ref. column includes references for experimental values }
\begin{tabular*}{\textwidth}{@{\extracolsep{\fill}}cccc}
 Molecule&True $D_0^0$ (eV) &Predicted $D_0^0$ (eV)& Ref.  \\
\hline
\\
RuC&6.34& 6.09 $\pm$ 1.28& \cite{39MoCRuCPdCNbCZrC} \\
\\
 MoC&5.01&5.85 $\pm$ 1.26&\cite{39MoCRuCPdCNbCZrC}, \cite{40MoC}\\
 \\
NbC&6.85&5.42 $\pm$ 1.17&\cite{45NbC}, \cite{39MoCRuCPdCNbCZrC} \\ 
\\
YC&4.29& 4.31 $\pm$ 1.34& \cite{65YC}, \cite{66YC} \\ 
\\
ZnBr&2.45& 3.86 $\pm$ 0.63&\cite{61ZnBr} \\
\\ 
InBr&3.99& 4.05 $\pm$ 0.73&\cite{20InBr} \\
\\ 
SnI&2.89&3.33 $\pm$ 0.70&\cite{50PbISnI} \\
\label{table:do test}
\end{tabular*}
\end{center}
\end{table*}

\subsection{Testing ML models versus ab initio results}

To further assess the accuracy of our ML models regarding $R_e$ and $\omega_e$ we have exposed our models to molecules containing Fr. Indeed, our dataset does not contain any Fr-containing molecule, defining the most complicated scenario for our ML models. The results of r2 and w2 models in comparison with the state-of-the-art ab initio methods are shown in Table.~\ref{tablecomp}, where it is noticed that our ML predictions agree well with ab initio predictions. Furthermore, more data can quickly improve ML predictions, as presented in Figs.~\ref{fig:Re scatter} and \ref{We scatter}. Therefore, ML predictions can be competitive with ab initio quantum chemistry methods using a larger dataset.

\begin{table*}[h]
\begin{center}
\caption{$R_e$ and $\omega_e$ ML predictions for molecules not contemplated in the databse. The ab-initio results are taken from Ref.~\cite{ladjimi2023diatomic}.}
\begin{tabular*}{\textwidth}{@{\extracolsep{\fill}}ccccc}
 Molecule& Ab-inito $R_e$ (\AA) & r2 Predicted $R_e$ (\AA)&  Ab-inito $\omega_e$ (cm$^{-1}$) & w2 Predicted $omega_e$ (cm$^{-1}$) \\
\hline
LiFr& 3.691 & 3.709$\pm$ 0.123 & 180.7 & 198.9 $\pm$ 35.9 \\
 KFr& 4.284 & 4.483 $\pm$ 0.173 & 65.2 & 64.0  $\pm$ 16.2\\
RbFr& 4.429  & 4.389 $\pm$ 0.145 & 46.0 & 48.8  $\pm$ 10.4 \\ 
CsFr& 4.646 & 4.403 $\pm$ 0.221 & 37.7 & 42.7  $\pm$ 13.7 
\label{tablecomp}
\end{tabular*}
\end{center}
\end{table*}

\subsection{Predicting homonuclear spectroscopic properties from heteronuclear data}

\begin{figure*}[h]
\begin{center}
 \includegraphics[width=1\linewidth]{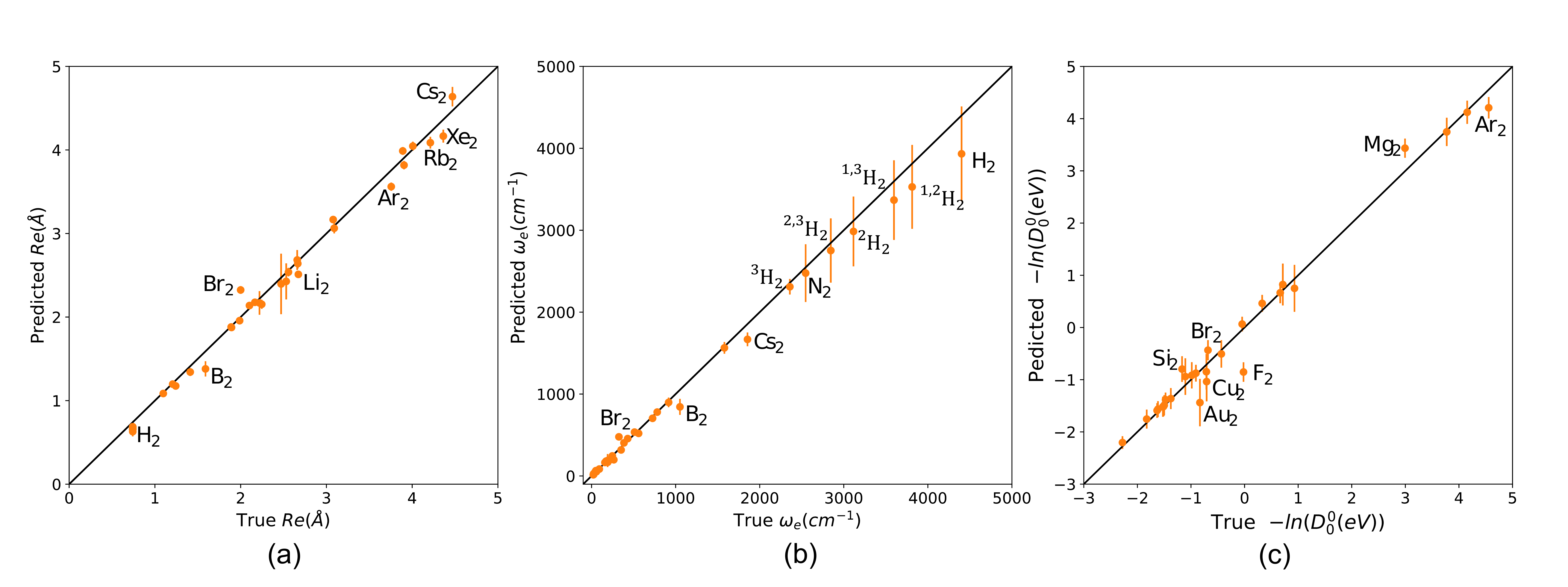}  
\caption{\label{fig:Homo from Hetro scatter}Scatter plots of models predicting $R_e$, $\omega_e$ and $-\ln(D_0^0)$ for homonuclear molecules from heteronuclear molecules data (a) r3 (b) w4 (c) d1. Points and error bars represent the predictive distribution means and standard deviations, respectively.}
\end{center}
\end{figure*}

To explore the capability of our models in predicting the spectroscopic properties of homonuclear molecules from spectroscopic and atomic information of heteronuclear molecules, we train our models for predicting $R_e$, $\omega_e$, and $D_0^0$ (r3, w4, and d1) using a special split. We fit the three models to heteronuclear data in $D_{\text{tv}}$ and then make predictions for the left-out homonuclear molecules. The performance of our models is displayed in Fig.~\ref{fig:Homo from Hetro scatter}, where we notice an outstanding performance. Only a few outliers are observed, showing a minimal deviation from the true line. In particular, we obtain MAEs of 0.08 \AA, 74 $\text{cm}^{-1}$, and 0.149 for models r4, w4, and d1, respectively. Hence, it is possible to predict the accurate spectroscopic properties of homonuclear molecules from heteronuclear data. Furthermore, our results indicate that expanding the data set by including homonuclear molecules yield high-performing models able to predict spectroscopic properties for both heteronuclear and homonuclear molecules.

\subsection{Towards a classification of diatomic molecules} \label{classification}

\begin{figure}[h]
\begin{center}
 \includegraphics[width=1\linewidth]{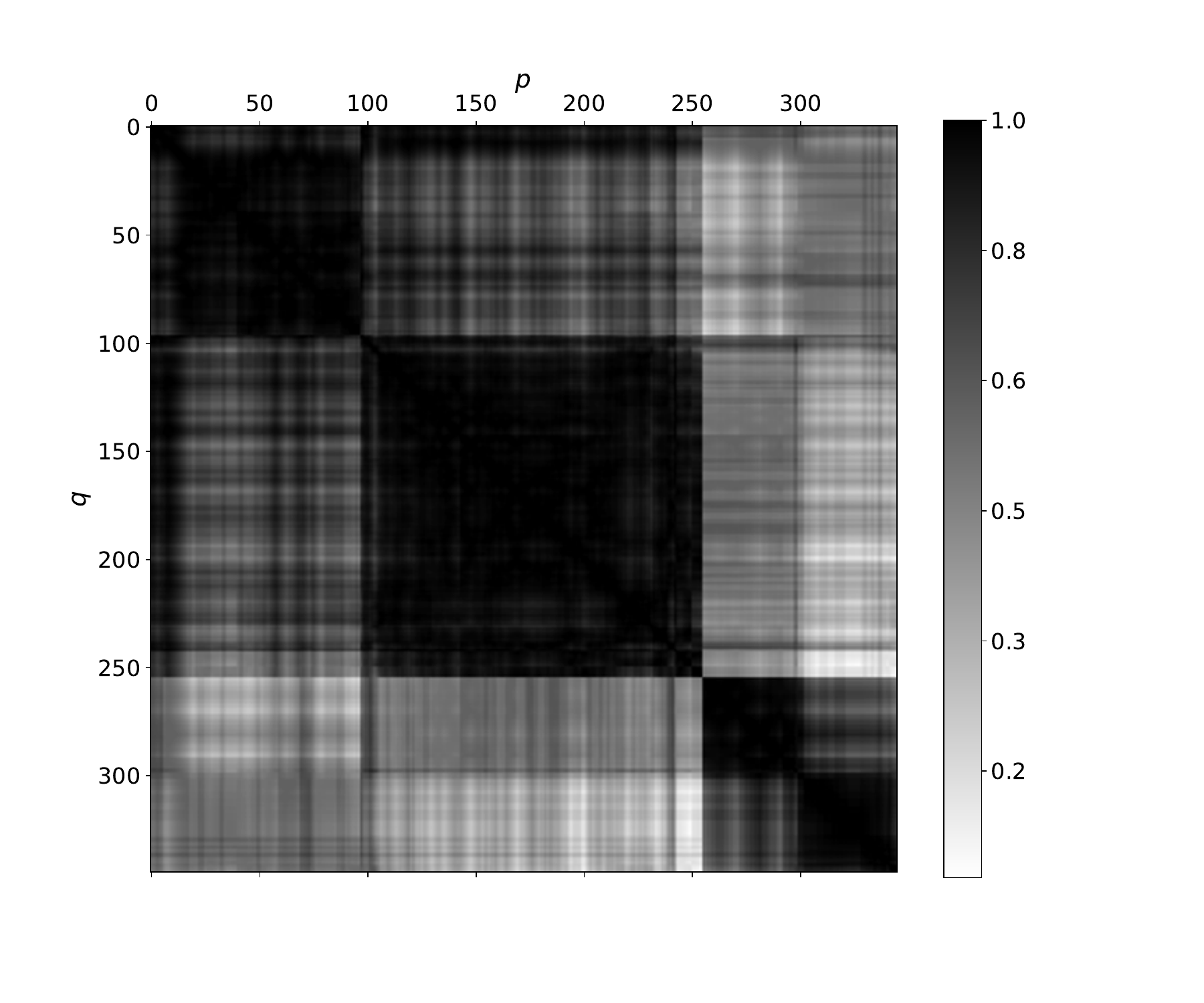}  
\caption{\label{fig:big Heat map}A Heat map quantifies the degree of similarity among molecules in the data set from 0 (white, not similar) to 1 (black, identical) on a grayscale. The Heat map was generated by finding the matrix element of a similarity matrix. Each matrix element quantifies the similarity between a pair of molecules $p$ (on the x-axis) and $q$ (on the y-axis) via Eq.\ref{eq10} with $\nu=3/2$ and parameters determined as described in the text.}
\end{center}
\end{figure}

For models r2, r3, w2, w3, and d1, we have achieved good results using a kernel common to all five models. That Mat\'ern kernel given by Eq.~(\ref{eq10}) with $\nu=3/2$, is a similarity measure. Therefore, it is possible to quantify the similarity between a pair of molecules denoted by $i=p,q$ giving their feature vector $\textbf{x}_i= \left(p_1^{i},g_1^{i}, p_2^{i}, g_2^{i}, \sqrt{\mu^{i}}\right)$. The models are fitted to the whole dataset to determine the parameters $(\sigma_n,l,\sigma_f)$. The kernel given by Eq.~(\ref{eq10}) with $\nu=3/2$ and the determined parameters can be used to form a similarity matrix. Each element in the similarity matrix quantifies the similarity between a pair of molecules in the dataset. Off-diagonal elements are calculated via Eq.~(\ref{eq10}) for $p \neq q$, with the diagonal representing the similarity of the molecules with themselves ($p=q$). A heat map representation of the similarity matrix is given in Fig.~\ref{fig:big Heat map}, while the degree of similarity from 0 to 1 is given over a greyscale as indicated by the color bar on the right side of the figure. 

\begin{figure*}[h]
\begin{center}
 \includegraphics[width=1\linewidth]{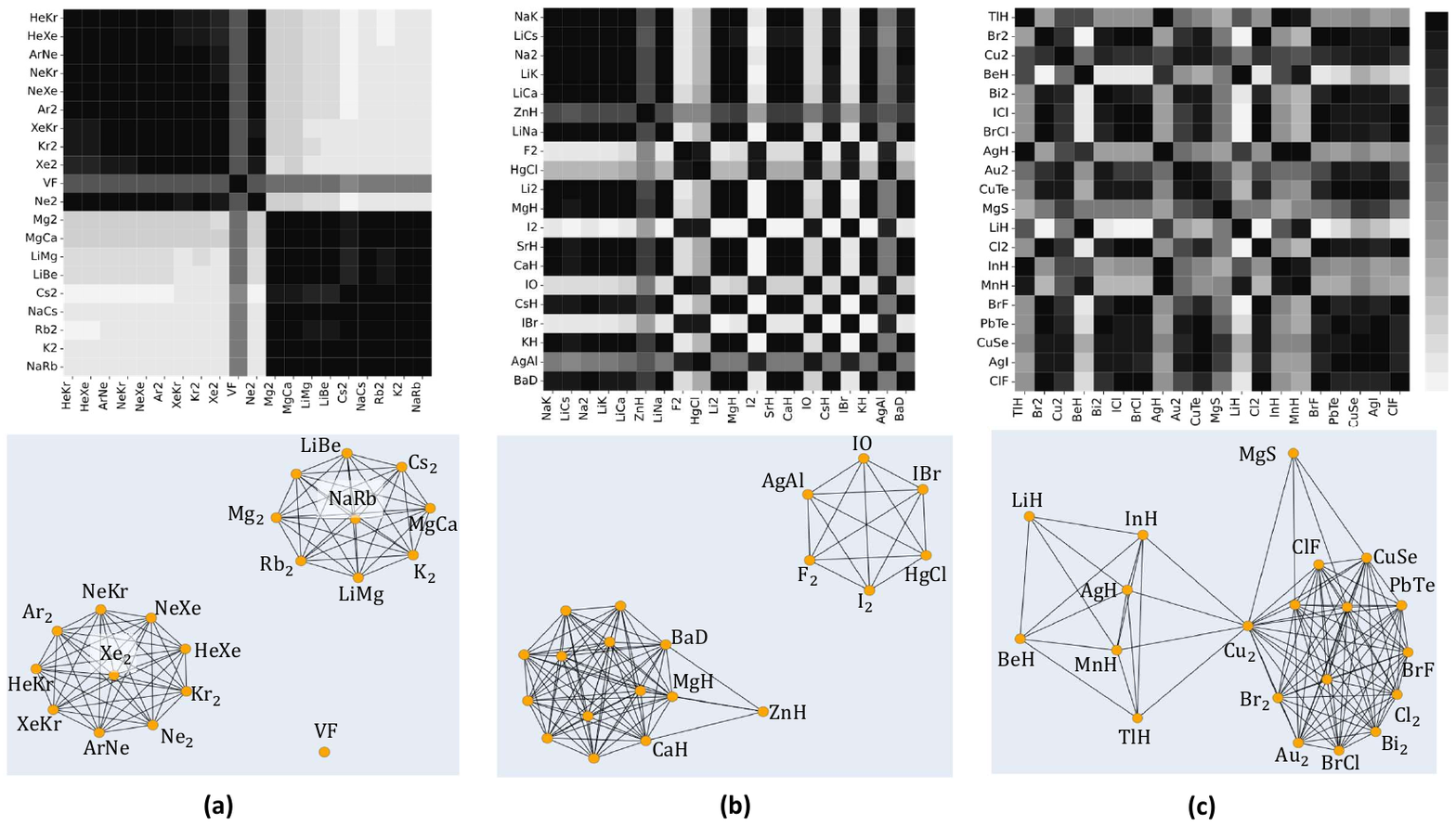}  
\caption{\label{fig:Heat maps} Heat maps representing similarities among subsets of molecules (upper row) and their corresponding network representation (lower row). The color bar (top right) quantifies the similarity between a pair of molecules from 0 (white, not similar) to 1 (black, identical) on a greyscale. The network representations show similarities above the 80\% (0.8) level. Each node represents a molecule. Short lines joining two nodes represent a high degree of similarity, while longer lines represent a lower degree of similarity above 80\%. No line at all indicates a lower degree of similarity below 80\%. }
\end{center}
\end{figure*}

To further explore the quantified similarity among molecules, we consider three subsets of molecules and show their heatmaps in the upper panels of Fig.~\ref{fig:Heat maps}. The lower panels of Fig.~\ref{fig:Heat maps} show the corresponding network representation of the similarity among these subsets of molecules. Black squares in the heat map plots of  Fig.~\ref{fig:Heat maps} indicate that a pair of molecules is highly similar, whereas white squares indicate 0\% similarity. The network representation represents each molecule as a node. The similarity between two molecules is diagrammatically shown with a line joining their corresponding nodes. The networks show similarities above the 80\% level. A line joins two nodes only if they are at least 80\% similar. The length of a joining line indicates the degree of similarity between a pair above the 80\% level. A short line indicates a high degree of similarity, and a long line indicates a lower degree of similarity.

From panel (a) of Fig.~\ref{fig:Heat maps}, we see noble gas dimers clustering around $\text{Xe}_2$, and alkali metals-alkaline earth metals cluster around $\text{NaRb}$. Both clusters are isolated from each other and VF, indicating a lower degree of similarities between these clusters and $\text{VF}$. A similar scenario is observed in panel (b), where alkaline earth metal hydrides cluster upon themselves with tight interconnections indicating high similarity. On the other hand, ZnH is remotely connected to the cluster, indicating a lower degree of similarity. The upper right cluster shows an interconnection among diatomic reactive nonmetals, including halides and oxygen; notably, ${\text{AgAl}}$ is connected to these molecules. Panel (c) displays a more involved clustering scheme involving transition metal hydrides (MnH and AgH), connected to a metalloid hydride (TlH and InH) and with a lower degree to alkaline earth metals hydrides (LiH and BeH). The right-hand side cluster consists of various transition metal diatomics, dihalides, and others, all closely related except for MgS. Note that all the molecules in the right-hand side cluster consist of atoms from the periodic table's right side. At the same time, MgS combine one atom from group 2 and one from group 16. Notably, homonuclear diatomic and heteronuclear molecules are firmly embedded within all the clusters, emphasizing the importance of including homonuclear data in our models.

Since only atomic properties are required to find elements of the matrix representation of the kernel, the similarity matrix can guide us in our data-gathering efforts. For example, we can determine which molecules can fill the gaps and connect clusters to build more robust models. More interestingly, we can systematically classify molecules based on the similarity matrix. Such classification would help develop potential energy surfaces (PES) for diatomic molecules. As pointed out by Newing, similar molecules will have similar potential energy surfaces, at least around $R_e$ \cite{newing1940xxx}.

\section{Summary and Conclusion}

In this work, first, we have extended the previous database of Liu et al.~\cite{liu2021relationship}, gathering ground state spectroscopic data of 86 homonuclear and heteronuclear molecules leading to a data set of 339 molecules. Next, the database has been used to train 9 ML models to predict the main spectroscopic constants: $R_e$, $\omega_e$, and $D_0^0$. These models can be categorized into three categories:

\begin{itemize}

\item Models in category (\textit{i}) only employ information from the periodic table and thus can predict spectroscopic properties of any combination of two elements. These models can be used to systematically classify molecules made up of any two elements in the periodic table (section \ref{classification}).  While spectroscopic data availability does not limit these models' ability to predict spectroscopic constants of any molecule, it affects the models' accuracy. These models are characterized by a relatively larger gap between validation and learning curves than models in categories (\textit{ii}) and (\textit{iii}). Thus, we would expect a better performance of category (\textit{i})  models upon training on larger datasets.



\item Models in category (\textit{ii}) use spectroscopic information only in their mean function but not in the kernel, and  are robust against noise in input variables. In this case,  since the mean function is a linear function, we can apply standard errors-in-variables methods \cite{van2013total}. This might be advantageous if we would like to use uncertain experimental data or predictions from (\textit{i}) models or {\it ab initio} methods to train our models.  

\item Models in category (\textit{iii}) include our most flexible, accurate, and consistent models(r4, w4). These models benefit from a high learning rate and a narrow gap between validation and learning curves. Apart from their outstanding performance, we can train these models using ground and excited states simultaneously since each state will be labeled by its spectroscopic constant values $R_e$ or $\omega_e$ along with other properties that define the molecule  \{$p_1$,$g_1$,$p_2$,$g_2$,$\mu^{1/2}$\}.      


\end{itemize} 

In summary, the newly developed models in this work showed an outstanding performance in all metrics in comparison to the previous ML models and other empirical and semiempirical models, with mean absolute errors ranging between 0.02~\AA~and 0.04~\AA~ for $R_e$ and 26 cm$^{-1}$ to 40  cm$^{-1}$ for $\omega_e$. We have been able to predict homonuclear spectroscopic properties with good accuracy upon training our models on heteronuclear molecules' data. Indeed, our models are almost as accurate as the state-of-the-art {\it ab inito} methods for diatomics~\cite{Liu2023,ladjimi2023diatomic}.

On the other hand, since we use the same kernel for all models under consideration, we are uniquely positioned to study a way to classify diatomic molecules beyond the traditional one based on the nature of the chemical bond. We expect such classification to enhance the performance and facilitate the development of ML models predicting spectroscopic and molecular properties of diatomic molecules. Further, the classification of diatomic molecules should help develop potential energy surfaces (PES). 


Finally, we have shown that for molecules with large ionic character and containing heavy atoms (e.g., LiCs, LiCa, AuF, and ZnBr), our predictions are comparable to DFT and even the state-of-the-art {\it ab initio} methods. Moreover, two of our models (r2 and w2) offer a promising opportunity to predict spectroscopic properties from atomic positions in the periodic table with high accuracy. This is a stepping stone towards closing the gap between atomic and molecular information; more spectroscopy data is required to do so. More extensive, open, and user-friendly data will help the field of data-driven science to help understand the chemical bonding and spectroscopy of small molecules. Indeed, that is something that we are currently working on in our group: we need more spectroscopic data in the big data era.   

\section*{Author Contributions}
X.L. helped with the database and the first ML models. M.A.E.I. gathered the data and performed the new ML models, whereas J.P.-R. envisioned the idea and supervised the project. M.A.E.I and J.P.-R. wrote the paper. 

\section*{Conflicts of interest}
There are no conflicts to declare.

\section*{Acknowledgements}
J.P.-R. acknowledges the funding of Simons Foundation and the Lorentz Center of the University of Leiden for organizing the workshop ``New directions in cold and ultracold molecules'', in which some part of this work was discussed. X.L. acknowledges the support or the the Deutsche Forschungsgemeinschaft (DFG – German Research Foundation) under the grant number PE 3477/2 - 493725479.  
M.A.E.I. acknowledges the funding and support of THE BINATIONAL FULBRIGHT COMMISSION and Assiut University in EGYPT.



\balance


\bibliography{rsc} 
\bibliographystyle{rsc} 

\end{document}